\documentclass[acmsmall]{acmart}

\usepackage{pifont}
\usepackage[ruled,vlined,linesnumbered]{algorithm2e}
\usepackage{array}
\usepackage[caption=false,font=footnotesize,labelfont=sf,textfont=sf]{subfig}
\usepackage{url}
\usepackage{booktabs}
\usepackage{epstopdf}
\usepackage{tabularx}
\usepackage{multirow}
\usepackage{footmisc}
\usepackage{arydshln}
\newcommand{\cmark}{\ding{51}}
\newcommand{\xmark}{\ding{55}}
\DeclareMathOperator*{\argmax}{arg\,max}
\usepackage{amsthm}
\theoremstyle{definition}
\newtheorem{definition}{Definition}[section]
\newcommand{\tabitem}{~~\llap{\textbullet}~~}

\setcopyright{acmcopyright}
\acmJournal{TIST}
\acmYear{2020} \acmVolume{1} \acmNumber{1} \acmArticle{1} \acmMonth{1} \acmPrice{15.00}

\begin{document}

\title{Latent Unexpected Recommendations}

\author{Pan Li}
\affiliation{\institution{New York University, Stern School of Business}}
\email{pli2@stern.nyu.edu}

\author{Alexander Tuzhilin}
\affiliation{\institution{New York University, Stern School of Business}}
\email{atuzhili@stern.nyu.edu}

\begin{abstract}
Unexpected recommender system constitutes an important tool to tackle the problem of filter bubbles and user boredom, which aims at providing unexpected and satisfying recommendations to target users at the same time. Previous unexpected recommendation methods only focus on the straightforward relations between current recommendations and user expectations by modeling unexpectedness in the feature space, thus resulting in the loss of accuracy measures in order to improve unexpectedness performance. Contrast to these prior models, we propose to model unexpectedness in the latent space of user and item embeddings, which allows to capture hidden and complex relations between new recommendations and historic purchases. In addition, we develop a novel Latent Closure (LC) method to construct hybrid utility function and provide unexpected recommendations based on the proposed model. Extensive experiments on three real-world datasets illustrate superiority of our proposed approach over the state-of-the-art unexpected recommendation models, which  leads to significant increase in unexpectedness measure without sacrificing any accuracy metric under all experimental settings in this paper.
\end{abstract}

\ccsdesc[500]{Information systems~Recommender systems}

\keywords{Unexpected Recommendation, Beyond-Accuracy Objectives, Latent Closure, Latent Embeddings, Latent Space}

\maketitle

\section{Introduction}
Recommender systems have been playing an important role in the process of information dissemination and online commerce, which assist users in filtering the best content while shaping their consumption behavior patterns at the same time. However, classical recommender systems are facing the problem of filter bubbles \cite{pariser2011filter,nguyen2014exploring}, which means that target users only get recommendations of their most familiar items, while losing reach to many other available items. They also lead to the problem of user boredom \cite{kapoor2015like,kapoor2015just}, which significantly deteriorates user satisfaction with recommender systems. For example, even a Harry Potter fan may feel unsatisfied if the system keeps recommending Harry Potter series all the time. 

To address these two problems, researchers have introduced recommendation objectives beyond accuracy, including unexpectedness, serendipity, novelty and diversity \cite{shani2011evaluating}, the goal of which is to provide novel, surprising and satisfying recommendations. Among them, unexpectedness is of particular interest for its close relation with user satisfaction and ability to improve recommendation performance \cite{adamopoulos2014discovering,adamopoulos2015unexpectedness}. Therefore, we focus on modeling unexpectedness and providing unexpected recommendations in this paper.

In prior literature, researchers have proposed to define unexpectedness in multiple ways, including deviations from primitive prediction results \cite{murakami2007metrics,ge2010beyond}, unexpected combination of feature patterns \cite{akiyama2010proposal} and feature distance from previous consumptions \cite{adamopoulos2015unexpectedness}. They subsequently provide unexpected recommendations based on these definitions and achieve significant performance improvements in terms of certain unexpectedness measures.

However as shown in the prior literature \cite{zolaktaf2018generic,zhou2010solving}, improvements in unexpectedness come at the cost of sacrificing accuracy measures, which severely limits practical use of unexpected recommendations since the major goal of recommender system is to enhance overall user satisfaction. This is the case for the following reasons. First, previous models only focus on the straightforward relations between current recommendation and user expectations by modeling unexpectedness in the \textit{feature space}, while not taking into account deep, complex and heterogeneous relations between users and items. Second, prior modeling of unexpectedness relies completely on the explicit user and item information, and may not work well in the case when the consumption records are sparse, noisy or even missing. And finally, the distance metric between discrete items, which is crucial for defining unexpectedness, is hard to formulate in the discrete feature space, and this may lead to unintentional biases in the estimation of user preferences. Therefore, prior unexpected recommendation models can be further improved, and this constitutes the main topic of this paper.

To address the aforementioned concerns, in this paper we propose to define unexpectedness in the \textit{latent space} containing latent embeddings of users and items, as opposed to the \textit{feature space} that only has the explicit information about them. Specifically, we propose a novel \textit{Latent Closure (LC)} method to model unexpectedness that:
\begin{itemize}
\item captures latent, complex and heterogeneous relations between users and items to effectively model the concept of unexpectedness.
\item provides unexpected recommendations without sacrificing any performance accuracy. 
\item efficiently computes unexpectedness for large-scale recommendation services.
\end{itemize}

The proposed unexpected recommendation model follows the following three-stage procedure. First, we map the features of users and items into the latent space and represent users and items as latent embeddings there. These embeddings are obtained using several state-of-the-art mapping approaches, including \textit{Heterogeneous Information Network Embeddings} (HINE) \cite{sun2013mining,shi2018heterogeneous,dong2017metapath2vec}, \textit{AutoEncoder} (AE) \cite{hinton2006reducing,sedhain2015autorec} and \textit{MultiModal Embeddings} (ME) \cite{Pan_2016_CVPR} methods. We subsequently utilize the concept of `'closure'' from differential geometry and formulate the definition of unexpectedness of a new item as the distance between the embedding of that item and the closure of all the previously consumed item embeddings. And finally, we combine this unexpectedness measure with the estimated rating of the item to construct the hybrid utility function for providing unexpected recommendations.

In this paper, we make the following contributions:

(1) We propose \textit{latent} modeling of unexpectedness. Although many papers have recently explored latent spaces for recommendation purposes, it is not clear how to do it for \textit{unexpected} recommendations, which constitutes the topic of this work.

(2) We construct hybrid utility function based on the proposed unexpectedness measure and provide unexpected recommendations accordingly. We also demonstrate that this approach would significantly outperform all other unexpected recommendation baselines considered in this paper.

(3) We conduct extensive experiments in multiple settings and show that it is indeed the latent modeling of unexpectedness that leads to significant increase in unexpectedness measures without sacrificing any accuracy performance. Thus, the proposed method helps users to break out of their filter bubbles without sacrificing recommendation performance. 

The rest of the paper is organized as follows. We discuss the related work in Section 2 and present our proposed latent modeling of unexpectedness in Section 3. The unexpected recommendation model is introduced in Section 4. Experimental design on three real-world datasets are described in Section 5 and the results as well as discussions are presented in Section 6. Finally, Section 7 summarizes our contributions and concludes the paper.

\section{Related Work}
In this section, we provide an overview on the related work covering three fields: beyond-accuracy metrics, unexpected recommendations and latent embeddings for recommendations. We highlight the importance of combining unexpected recommendations with latent modeling approaches to achieve superb recommendation performance.

\subsection{Beyond-Accuracy Metrics}
As researchers have pointed out, accuracy is not the only important objective of recommendations \cite{mcnee2006being}, while other beyond-accuracy metrics should also be taken into account, including unexpectedness, serendipity, novelty, diversity, coverage and so on \cite{ge2010beyond,kaminskas2016diversity}. Note that, these metrics are closely related to each other, but still different in terms of definition and formulation. Therefore, prior literature have proposed multiple recommendation models to optimize each of these metrics separately. 

Serendipity measures the positive emotional response of the user about a previously unknown item and indicates how surprising these recommendations are to the target users\cite{shani2011evaluating,chen2019serendipity}. Representative methods to improve serendipity performance include Serendipitous Personalized Ranking (SPR) \cite{lu2012serendipitous} that extends traditional personalized ranking methods by considering serendipity information in the AUC optimization process; and Auralist\cite{zhang2012auralist} that utilizes the topic modeling approach to capture serendipity information and provide serendipitous recommendations accordingly.

Novelty measures the percentage of new recommendations that the users have not seen before or known about \cite{mcnee2006being}. It is computed as the percentage of unknown items in the recommendations. Researchers have proposed multiple methods to improve novelty measure in recommendations, including clustering of long-tail items \cite{park2008long}, innovation diffusion \cite{ishikawa2008long}, graph-based algorithms \cite{shi2013trading} and ranking models \cite{wasilewski2019bayesian,oh2011novel}.

Diversity measures the variety of items in a recommendation list, which is commonly modeled as the aggregate pairwise similarity of recommended items \cite{ziegler2005improving}. Typically models to improve diversity of recommendations include Determinantal Point Process (DPP) \cite{gartrell2017low,chen2018fast} that proposes a novel algorithm to greatly accelerate the greedy MAP inference and provide diversified recommendation accordingly; Greedy Re-ranking methods \cite{ziegler2005improving,smyth2001similarity,kelly2006enhancing,vargas2014coverage,barraza2017exploration} that provide diversified recommendations based on the combination of the item’s relevance and its average distance to items already in the recommended list; and also Latent Factor models to optimize diversity measures \cite{shi2012adaptive,hurley2013personalised,su2013set}

Coverage measures the degree to which recommendations cover the set of available items \cite{ge2010beyond,herlocker2004evaluating,adomavicius2011improving}. To improve coverage measure, researchers propose to use coverage optimization \cite{adomavicius2011improving,adomavicius2011maximizing} and popularity reduction methods \cite{vargas2011rank} to balance between relevance and coverage objectives \cite{wu2016relevance}. 

Over all beyond-accuracy metrics, in this paper we only focus on the unexpectedness measure and aim at providing unexpected recommendations for its close relation with user satisfaction and ability to improve recommendation performance \cite{adamopoulos2014discovering,adamopoulos2015unexpectedness}. Moreover, the proposed unexpected recommendation algorithm is capable of improving serendipity and diversity measures as well, as shown in our experiment results.

\subsection{Unexpectedness in Recommendations}
Different from other beyond-accuracy metrics, unexpectedness measures those recommendations that are not included in user expectations and depart from what they would expect from the recommender system. Researchers have shown the importance of incorporating unexpectedness in recommendations, which could overcome the overspecialization problem \cite{adamopoulos2015unexpectedness,iaquinta2010can}, broaden user preferences \cite{herlocker2004evaluating,zhang2012auralist,zheng2015unexpectedness} and increase user satisfaction \cite{adamopoulos2015unexpectedness,zhang2012auralist,lu2012serendipitous}. Unexpectedness captures the deviation of a particular recommender system from the results obtained from other primitive prediction models \cite{murakami2007metrics,ge2010beyond,akiyama2010proposal}, and also the deviation from user expectations \cite{adamopoulos2015unexpectedness,li2019latent3}.To improve unexpectedness measure in the final recommendations, existing models can be classified into three categories: rule-based approaches, model-based approaches and utility-based approaches, as we show in Table \ref{classification}. 

Rule-based approaches typically involve pre-definition of a set of rules or recommendation strategies for unexpected recommendations, including partial similarity \cite{kamahara2005community}, k-furthest-neighbor \cite{said2012increasing} and graph-based approaches \cite{taramigkou2013escape,lee2015escaping}. Rule-based approaches are generally simple to implement and easy to put into actual practice, as most of the approaches incorporate unexpectedness into the classical models instead of starting from scratch. Besides, rule-based approaches allow for more control in the model, as the rules and recommendation strategies are often explicitly specified by the designers. It also improves the explanability and interpretability of the proposed unexpected recommendation model. However, they require pre-defined strategies to be set prior to recommendations. Also, scalability is a big concern for the usage of rule-based methods. In addition, these models typically lack of generalizability for they focus only on specific domains and specific applications.

Model-based approaches aim to improve novelty and unexpectedness of the recommended items by proposing new models and data structures that go beyond the traditional collaborative filtering paradigm. Representative models that optimize the unexpectedness objective include personalized ranking \cite{wasilewski2019bayesian}, innovator identification \cite{kawamae2009personalized,kawamae2010serendipitous} and transition cost graph \cite{shi2013trading}. Model-based approaches are backed with mathematical foundations that guarantee either convergence or stability of the learning process, thus making them robust to different settings and with greater potential of generalizablity. However, they are often hard to interpret, for there is no natural way to transfer mathematical formulations into explicit rules or recommendation strategies. Therefore, it is relatively hard to control the degree of unexpectedness that we aim to incorporate into the recommendation model. And finally, model-based approaches might not take full advantages of all available information due to the restrictions of specific model form.

Utility-based approaches involve the construction of a hybrid utility function as the combination of estimated relevance and degree of unexpectedness. Researchers in \cite{weng2007improving,iaquinta2008introducing,hijikata2009discovery} have followed this direction of research. Specifically, \cite{adamopoulos2015unexpectedness} proposed to include user expectation into the hybrid utility function and achieves state-of-the-art unexpected recommendation performance. Utility-based methods allow for more control of the recommendation strategy, and it is easier to implement and put into practice as well. Especially, the construction of unexpectedness do not depend on the estimation of user preferences towards the candidate item, thus making it model-agnostic. On the other hand, the unexpected hyperparameter plays an important role in determining the recommendation performance of the hybrid-based model, thus requiring proper hyperparameter optimization.

One important limitation of all prior unexpected recommendation models lies in that they only focus on the straightforward relations between users and items and define unexpectedness in the feature space, without taking into account the deep, complex interactions underlying their feature information. Therefore, previous unexpected recommendations might not reach the optimal recommendation performance, as discussed in \cite{yu2013recommendation,shi2018heterogeneous,yu2014personalized}. In addition, they are facing the trade-off dilemma between optimizing the accuracy and unexpectedness objectives. To address these limitations, in this paper we propose to define unexpectedness instead in the latent space, thus obtaining significant improvements over previous models.

\begin{table*}
\centering
\resizebox{\textwidth}{!}{
\begin{tabular}{|l|l|l|l|} \hline
Model & Literature & Strength & Weakness \\ \hline
Rule-Based Approaches & \cite{said2012increasing}, \cite{chiu2011social}, \cite{kamahara2005community}, & \tabitem Easy to implement & \tabitem Require pre-defined rules \\
\tabitem K-Furthest Neighbor & \cite{lee2015escaping}, \cite{taramigkou2013escape} & \tabitem Allow for model control & \tabitem Lack of scalability \\
\tabitem Frequency Discount &  & \tabitem Improves interpretability & \tabitem Lack of generalizability \\
\tabitem Taxonomy-Based Similarity & & & \\ 
\tabitem Partial Similarity & & & \\ 
\tabitem Social Network & & & \\ 
\tabitem Graph Theory & & & \\ \hline
Model-Based Approaches & \cite{kawamae2009personalized}, \cite{kawamae2010serendipitous}, \cite{lu2012serendipitous}, & \tabitem Robust and generalizable & \tabitem Lack of interpretability \\ 
\tabitem Matrix Factorization & \cite{shi2013trading}, \cite{wasilewski2019bayesian} & \tabitem Mathematical foundation & \tabitem Restricted model control \\ 
\tabitem Learning to Rank & & \tabitem Efficient optimization & \tabitem Limited model input \\
\tabitem Re-Ranking & & & \\
\tabitem Clustering & & & \\ 
\tabitem Graph Theory & & & \\ \hline
Utility-Based Approaches & \cite{weng2007improving}, \cite{iaquinta2008introducing}, \cite{hijikata2009discovery}, & \tabitem Balance between objectives & \tabitem Require hyperparameter optimization \\
\tabitem Weighted Sum Model & \cite{zhang2012auralist}, \cite{adamopoulos2015unexpectedness}, \cite{li2019latent3} & \tabitem Allow for model control & \tabitem Explicit information only \\
\tabitem Weighted Product Model & & \tabitem Model-agnostic & \\
\tabitem Probabilistic Model & & & \\
\tabitem Neural Network Model & & & \\ \hline
\end{tabular}
}
\newline
\caption{Classification of Unexpected Recommendation Research}
\label{classification}
\end{table*}

\subsection{Latent Embeddings for Recommendation}
Another body of related work is around embedding approaches that effectively map users and items into the latent space and extract their deep, complex and heterogeneous relations between each other. Specifically, different embedding methods fit for different recommendation applications. In the case where heterogeneous feature data is available, Heterogeneous Information Network Embedding approach (HINE) \cite{shi2017survey,shi2018heterogeneous,dong2017metapath2vec} utilizes the data structure of heterogeneous information network (HIN) to extract complex heterogeneous relations between user and item features and thus provide better recommendations to the target users. In the case where rich interactions between users and items are available, AutoEncoding (AE) approach \cite{rumelhart1985learning,he2017neural,hinton2006reducing,sedhain2015autorec,li2019latent2,li2020ddtcdr} utilizes deep neural network (DNN) techniques and obtain the semantic-aware representations of users and items as embeddings in the latent space to model their relationship and provide recommendations accordingly. Finally in the case where multimodal dataset is available, researchers propose to use Multimodal Embedding (ME) approach \cite{Pan_2016_CVPR} to combine information from different sources and obtain superb recommendation performance.

Compared with classical approaches, latent embedding methods have several important advantages that enable recommender systems to provide more satisfying recommendations \cite{zhang2017deep,lin2005semantic}, as discussed in Section 1. Therefore, in this paper we provide the definition of unexpectedness utilizing these latent embedding methods, which contributes to the strong recommendation performance.

\section{Latent Modeling of Unexpectedness}
In this section, we introduce the proposed latent modeling of unexpectedness. We compare the new definition with feature-based definitions and illustrate superiority and benefits of the proposed approach.

\subsection{Latent Space}
As introduced in prior literature \cite{murakami2007metrics,ge2010beyond,adamopoulos2015unexpectedness}, an important component for modeling unexpectedness is the \textit{expected set}, which contains previous consumptions of the user. The idea is that, users should have no unexpectedness towards those recommended items that they have purchased before or very similar to their purchases, for they understand that typical recommender systems collect their historic behaviors and thus provide similar recommendations based on these records. 

To construct the user expectations, \cite{adamopoulos2015unexpectedness} propose to form the expected set by taking into account explicit feature information of users and items. For example in the book recommendation, the expected set is constructed based on the features of alternative editions, in the same series, with same subjects and classifications, with the same tags, and so on. Unexpectedness is subsequently defined by a positive, unbounded function of the distance of the recommended item from the set of expected items.

However, this definition only focuses on the straightforward relations between users and items, but fall short of addressing deeper correlations beyond the explicit feature information. For example, if a certain user has been a frequent consumer of McDonald and Carl's Jr, then the recommendation of Burger King might not be unexpected to that user, although these restaurants belong to different franchise and offer different menus, as shown in their feature information.

Besides, feature-based modeling of unexpectedness typically assumes the same importance for each feature during the calculation of unexpectedness, while in reality it is not necessarily the case. A natural example is that for music recommendations, genre information plays a more important role in determining the degree of unexpectedness than profile information, such as time of release. In addition, the distance function is also hard to define in the discrete feature space.

Therefore, in this paper, we propose to construct the expected set in the latent space by taking the closure of item embeddings. Unexpectedness is subsequently defined as the distance between the new item embedding and the closure of the expected set in the latent space. Comparing with feature-based definitions, latent modeling of unexpectedness obtains several important advantages, as discussed in Section 1. Especially, we point out that the proposed Latent Closure (LC) model is capable of utilizing richer information of user reviews and multi-modal data to determine the degree of unexpectedness, as previous models typically do not take these information into account, as shown in Table \ref{compare}. These benefits are also supported by strong experiment results.

\begin{table}[ht]
\centering
\resizebox{0.7\columnwidth}{!}{
\begin{tabular}{|c|c|cccc|} \hline
& Latent Modeling &\multicolumn{4}{c|}{Feature Modeling} \\ \hline
Algorithms                & LC     & SPR      & Auralist & HOM-LIN & DPP  \\ \hline
Latent Embeddings   & \cmark  & \xmark & \xmark & \xmark & \xmark \\
Explicit Features   & \cmark  & \xmark & \xmark & \cmark & \xmark \\
User Reviews           & \cmark  & \xmark & \cmark & \xmark & \xmark \\
Pre-Defined Rules   & \cmark   & \xmark & \cmark & \cmark & \xmark \\
Past Transactions     & \cmark   & \cmark & \cmark & \cmark & \cmark \\
User Ratings                     & \cmark  & \cmark & \cmark & \cmark & \cmark \\ \hline
\end{tabular}
}
\caption{Comparison of Unexpected Recommendation Methods}
\label{compare}
\end{table}

In the next section, we will introduce the idea of latent closure and how to construct user expectations based on the proposed latent closure method.

\subsection{Latent Closure}
As discussed in the previous section, we propose to compute user expectations in the latent space rather than in the original feature space. In addition, we point out that the modeling of user expectations should go beyond the direct aggregation of previous consumptions, and should also take into account those items that are similar to the consumed items, while similarities between items are captured by the Euclidean distance in the latent space. Therefore, it is natural to take the `'closure'' of all consumed item embeddings to model the expected set, as opposed to using individual item embedding in the latent space.

According to mathematical theories in differential geometry \cite{helgason2001differential}, there are three common geometric structures in high-dimensional latent spaces that can be naturally extended to modeling the closure of latent embeddings, namely Hypersphere, Hypercube and Convex Hull. The particular choice of latent closure depends on the assumption we make towards the relations between users and items in the latent space.

\begin{itemize}
\item \textbf{Latent HyperSphere (LHS)} The hypersphere in the $R^{n}$ space is defined as the set of n-tuples points ($x_{1},x_{2},\cdots.x_{n}$) such that $x_{1}^{2}+x_{2}^{2}+\cdots+x_{n}^{2}=r^{2}$ where r is the radius of the hypersphere. Under this definition, we assume that the expected set of items for each user grows homogeneously in all directions in the latent space.
\item \textbf{Latent HyperCube (LHC)} The hypercube is a closed, compact, convex figure, whose 1-skeleton consists of groups of opposite parallel line segments aligned in each of the space's dimensions, perpendicular to each other and of the same length. Under this definition, we assume that the expected set of each user grows homogeneously in the n perpendicular directions.
\item \textbf{Latent Convex Hull (LCH)} The convex hull of a set of points $X$ in the Euclidean space is the smallest convex set that contains all points in $X$. Under this definition, we assume that the expected set maintains its convexity in the growing process. In addition, if we construct the expected set as the convex hull of consumed item embeddings, the convexity property will guarantee the feasibility of the recommendation as an optimization problem given by the Slater's Condition \cite{slater2014lagrange}.
\end{itemize}

\begin{figure*}
\centering
\subfloat[Latent Convex Hull]{\includegraphics[width=.33\textwidth]{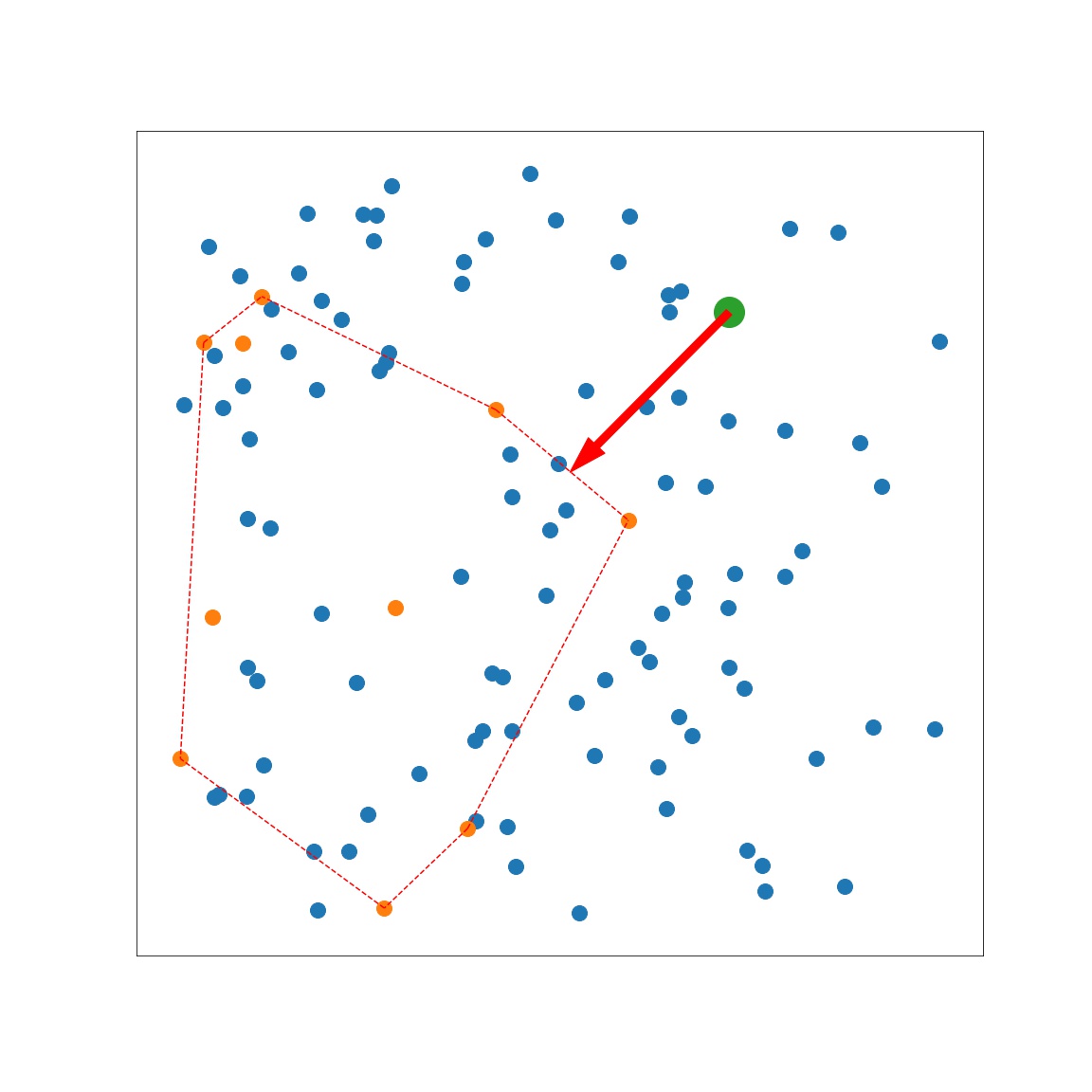}\label{fig1:sub1}}
\hfil
\subfloat[Latent Hypersphere]{\includegraphics[width=.33\textwidth]{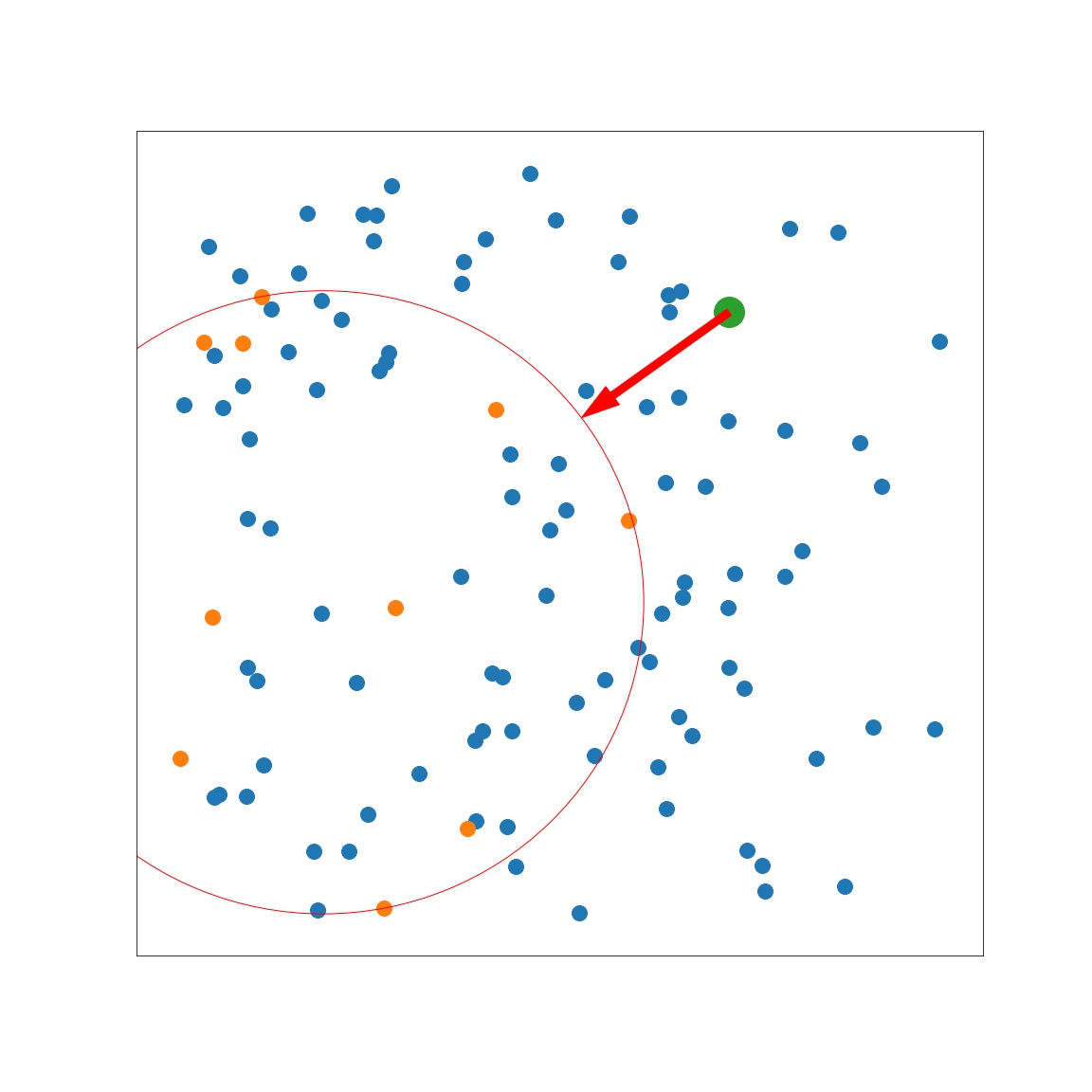}\label{fig1:sub2}}
\hfil
\subfloat[Latent Hypercube]{\includegraphics[width=.33\textwidth]{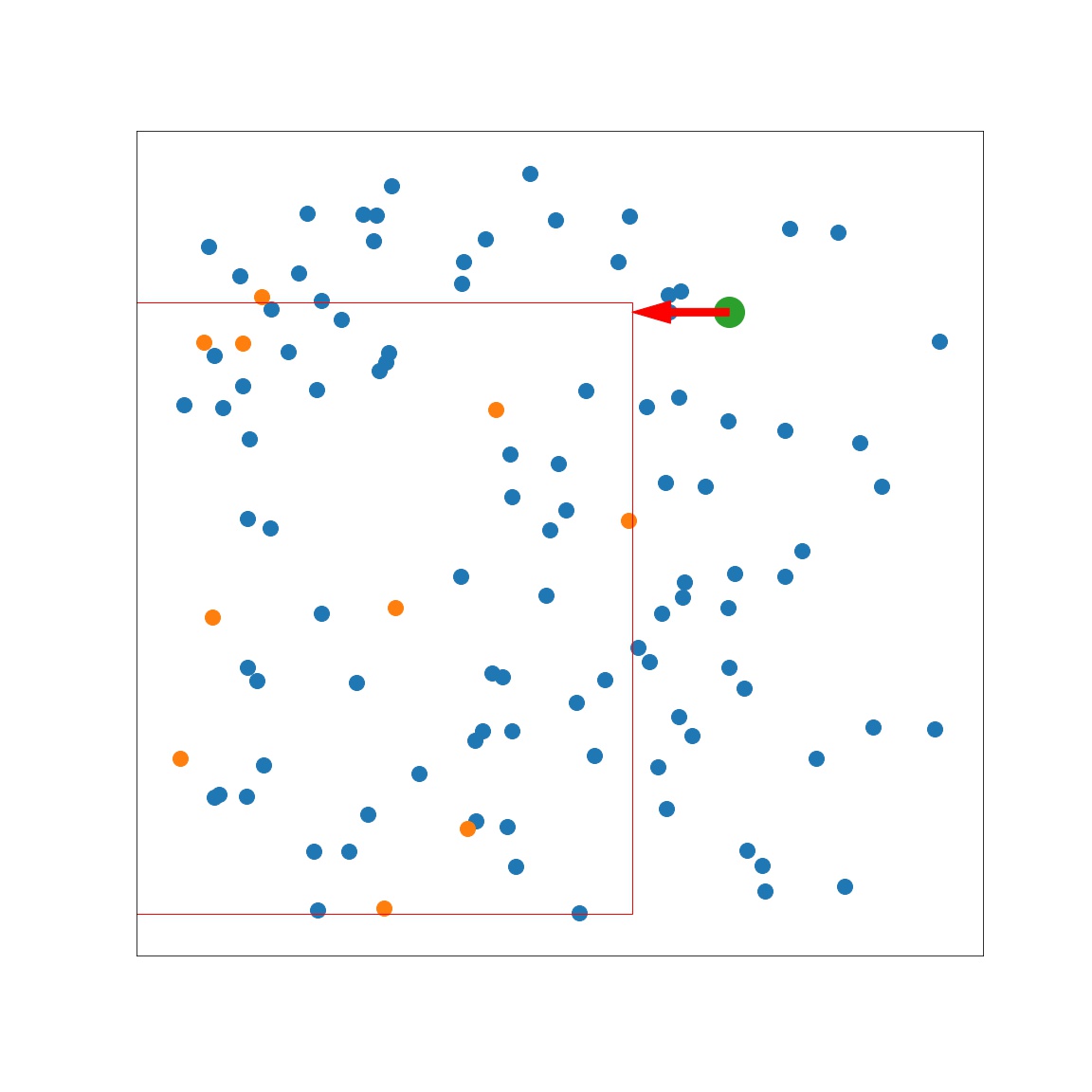}\label{fig1:sub3}}
\caption{Visualization of Latent Closure and the Unexpectedness. Blue points stand for all the available items; Orange points represent the consumed items; Green point refers to the newly recommended item. We define unexpectedness as the distance between the new item and the latent closure generated by all consumed items.}
\label{definition}
\end{figure*}

We visualize the definition of unexpectedness based on these geometric structures in Figure \ref{fig1:sub1}, \ref{fig1:sub2} and \ref{fig1:sub3}. These latent closure approaches capture latent semantic interactions between users and items and construct the expected set for each user accordingly. Compared to feature-based definitions \cite{adamopoulos2015unexpectedness}, latent closures utilize richer information including user and item features to model user expectations more precisely. The process for finding closures in high-dimensional latent spaces is not significantly different from the process in the 2-dimensional space. For LHS and LHC, we only need to find the furthest two points in the latent space to identify the centroid of the latent closure. For LCH, we follow the QuickHull algorithm \cite{barber1996quickhull} to identify the latent structure. Experiment results show that all three geometric structures consistently obtain significant improvements over baseline models, while no structure dominates the other two.

To sum up, in this paper we utilize the latent closure method to model unexpectedness in the latent space. We hereby propose the following definition of unexpectedness:
\theoremstyle{definition}
\begin{definition}{}
\textbf{Unexpectedness} of a new item as the distance between the embedding of that item and the closure of all previously consumed item embeddings. 
\end{definition}

In the next section, we will discuss the specific techniques for obtaining latent embeddings and methods to provide unexpected recommendations accordingly.

\section{Unexpected Recommendation Model}
\subsection{Latent Embeddings}
To effectively model unexpectedness in the latent space and demonstrate the robustness of the proposed model, we utilize three state-of-the-art latent embedding approaches, namely HINE, AE and ME to map users and items into the latent space and calculate the unexpectedness subsequently. 
\subsubsection{Heterogeneous Information Network Embeddings (HINE)}
To capture the complex and multi-dimensional relations in the data record, Heterogeneous Information Network (HIN) \cite{sun2013mining} has become an effective data structure for recommendations, which models multiple types of objects and multiple types of links in one single network. It includes users, items, transactions, ratings, entities extracted from reviews and the feature information. We link the associated entities with corresponding users and items in the network and utilize meta-path embedding approach \cite{dong2017metapath2vec} to obtain node embeddings.

We denote the heterogeneous network as $G=(V,E,T)$, in which each node $v$ and each link $e$ are assigned with specific type $T_{v}$ and $T_{e}$.  To effectively learn node representations we enable the skip-gram mechanism to maximize the probability of each context node $c_{t}$ within the neighbors of $v$, denoted as $N_{t}(v)$, where we add the subscript $t$ ($t \in T_{v}$) to limit the node to a specific type:
\begin{equation}
\argmax_{\theta}\sum_{v \in V}\sum_{t \in T_{v}}\sum_{c_{t} \in N_{t}(v)}log P(c_{t}|v;\theta)
\end{equation}
Thus, it is important to calculate $P(c_{t}|v;\theta )$, which represents the conditional probability of context node $c_{t}$ given node $v$. Therefore, we follow \cite{grover2016node2vec} and revise the network embedding model accordingly for dealing with heterogeneous information network. Specifically, we propose to use heterogeneous random walk to generate paths of multiple types of nodes in the network. Given a heterogeneous information network $G=(V,E,T)$, the metapath of the network is generated in the form of $V_{1} \xrightarrow{R_{1}} V_{2} \xrightarrow{R_{2}} V_{3} \cdots V_{n} $ wherein $R = R_{1} \circ R_{2} \circ \cdots R_{n} $ defines the composite relations between the start and the end of the heterogeneous random walk. The transition probability within each random walk between two nodes is defined as follows:
\begin{equation}
p(V_{t+1}| V_{t}) = \begin{cases}
    \frac{C(T_{V_{t}},T_{V_{t+1}})}{|N_{t+1}(V_{t})|},& (V_{t}, V_{t+1}) \in E \\
    0,              & (V_{t}, V_{t+1}) \notin E
\end{cases}
\end{equation}
where $C(T_{V_{t}},T_{V_{t+1}})$ stands for the transition coefficient between the type of node $V_{t}$ and the type of node $V_{t+1}$. We have 6 different transition coefficients that correspond to 6 different relations in the network $C_{UU}, C_{UE}, C_{UI}, C_{EI}, C_{EE}$ and $C_{II}$. (U:User, I:Item, E:Entity/Feature) $|N_{t+1}(V_{t})|$ stands for the number of nodes of type $V_{t+1}$ in the neighborhood of $V_{t}$.  We apply heterogeneous random walk iteratively to each node and generate the collection of meta-path sequences. The user and item embeddings are therefore obtained through the aforementioned skip-gram mechanism.

\begin{figure}
\centering
\includegraphics[width=.9\textwidth]{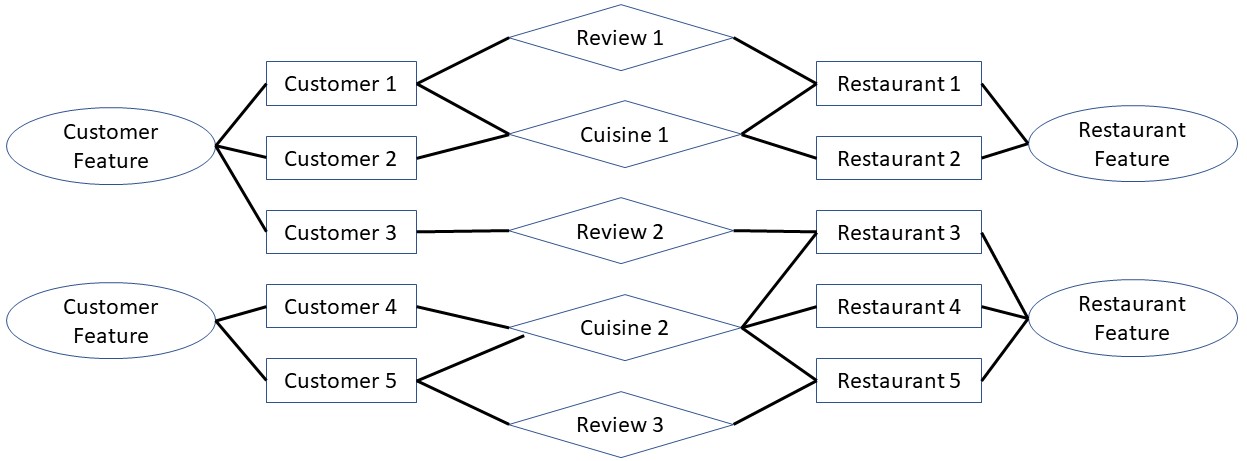}
\caption{Heterogeneous Information Network Embedding Method}
\label{multimodal}
\end{figure}

\subsubsection{AutoEncoder (AE)}
Apart from modeling interactions between users and items through HIN, AutoEncoder (AE) approach also constitutes an important tool to learn the latent representations of user and item features and transform discrete feature vectors into continuous feature embeddings. 

We denote the feature information for user $a$ as $u_{a} = \{u_{a_{1}},u_{a_{2}},\cdots,u_{a_{m}}\}$ and the feature information for item $b$ as $i_{b} = \{i_{b_{1}},i_{b_{2}},\cdots,i_{b_{n}}\}$, where $m$ and $n$ stand for the dimensionality of user and item feature vectors respectively. The goal is to train two separate neural networks: encoder that maps feature vectors into latent embeddings, and decoder that reconstructs feature vectors from latent embeddings. Due to effectiveness and efficiency of the training process, we formulate both the encoder and the decoder as multi-layer perceptron (MLP). MLP learns the hidden representations using the following equations:
\begin{equation}
y_{a} = \Phi(u_{a}), y_{b} = \Phi(i_{b})
\end{equation}
where $y_{a}, y_{b}$ represents the latent embeddings and $\Phi$ stands for the fully connected layer with activation functions. We apply another layer of fully connected network for reconstruction and optimization. Note that, in this step we train the global autoencoder for users and items in the entire dataset simultaneously to obtain the hidden representation.

\begin{figure}
\centering
\includegraphics[width=.5\textwidth]{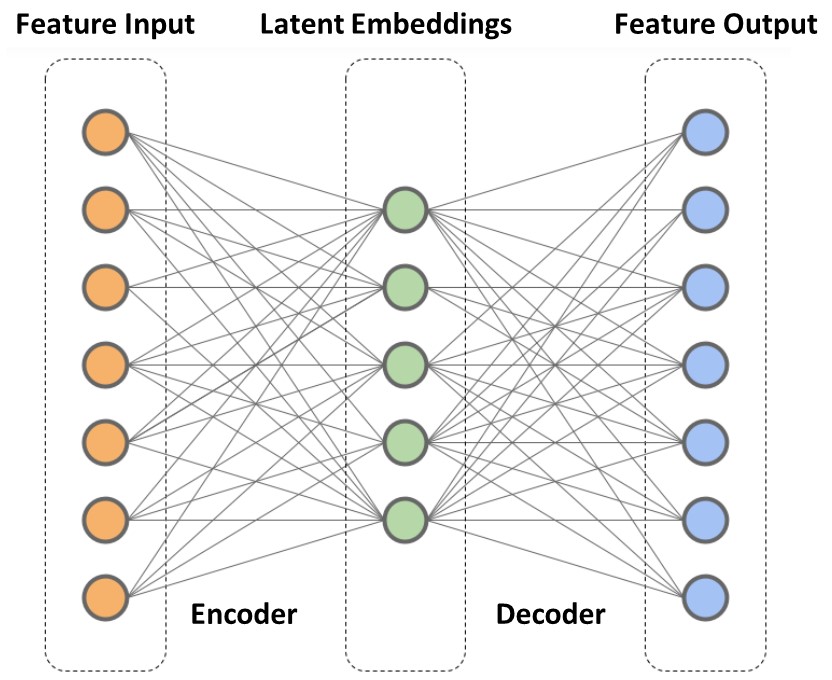}
\caption{AutoEncoder Embedding Method}
\label{multimodal}
\end{figure}

\subsubsection{Multimodal Embeddings (ME)}
In addition to the aforementioned approaches, when dealing with datasets that include multiple modalities, such as movie and video data (which are usually associated with images and subtitles), multimodal embeddings \cite{Pan_2016_CVPR,wei2019mmgcn} constitute an efficient tool to combine the information from different sources.

Specifically, in the video recommendation task, we illustrate the model for obtaining video embeddings in Figure \ref{multimodal}. First, we initialize the embeddings for text, audio and image data through Fully Convolutional Network (FCN) with L2-Norm as regularization term. For the text data, we use the average pooling technique as a special treatment to obtain the semantic information as the average of word embeddings. Then we concatenate these embeddings and apply another layer of Fully Convolutional Network to obtain multimodal embeddings for the input video that captures joint information of subtitles, sound and graphics.

\begin{figure}
\centering
\includegraphics[width=.7\textwidth]{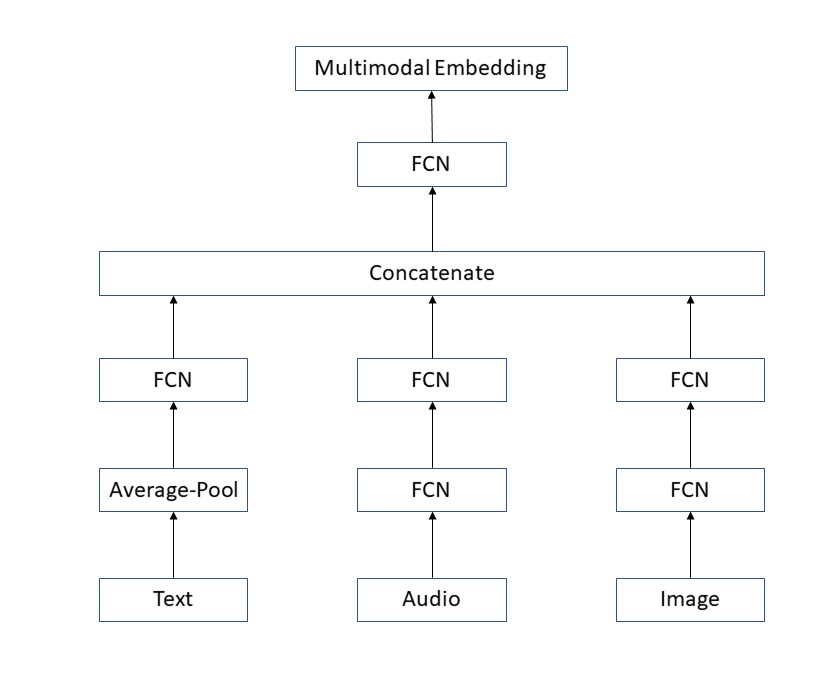}
\caption{Multimodal Embedding Method}
\label{multimodal}
\end{figure}

\subsection{Hybrid Utility Function}
Based on the latent embedding approaches introduced in the previous section, we map the users and items into the continuous latent space and model the expected set for each user as the latent closure of item embeddings. Specifically, we feed the user and item features as input into the latent embedding models and obtain their latent representations. We subsequently formulate the unexpectedness as the distance between the embedding of new item and the latent expected set as
\begin{equation}
U_{u,i} = d(i;LC(N_{i})) 
\end{equation}
where $N_{i} = (i_{1}, i_{2}, \cdots, i_{n})$ contains the embeddings of all consumed items. This unexpectedness metric is well defined as the minimal distance from the new item to the boundaries of the closure in the latent space. We then perform the unexpected recommendation based on the hybrid utility function:
\begin{equation} 
Utility_{u,i}= EstRating_{u,i} + \alpha *Unexp_{u,i}
\end{equation}
which incorporates the linear combination of estimated ratings and unexpectedness. The key idea lies in that, instead of recommending the similar items that the users are very familiar with as the classical recommenders do, we recommend unexpected and useful items to the users that they might have not thought about, but indeed fit well to their satisfactions. The two adversarial forces of accuracy and unexpectedness work together to get the optimal recommendation and thus obtain the best recommendation performance and user satisfaction. We present the entire framework in Algorithm \ref{algorithm}.

\begin{algorithm}
\SetArgSty{textnormal}
\SetAlgoLined
\KwData{Users; Items; Historic Actions; Other feature information}
\KwResult{List of Recommended Items}
Map users and items into the latent space\;
\For{each user $u$ in Users}{
\For{each item $i$ in Items}{
$Unexp_(u, i) = d(i;LC(N_{i}))$\;
$Utility_{u,i}= EstRating_{u,i} + \alpha*Unexp_{u,i}$}
Recommend Top-N(Utility)\;}
\caption{Latent Unexpected Recommendation}
\label{algorithm}
\end{algorithm}

\section{Experiments}
To validate the performance of our approach, we conduct extensive experiments on three large-scale real-world applications and compare the results of our model with the state-of-the-art baselines. The experimental setup is introduced in this section. Specifically, we design the experiments to address the following research questions:

\textbf{RQ1}: \textit{How does the proposed model perform compared to baseline unexpected recommendation models?}

\textbf{RQ2}: \textit{Can we achieve significant improvements in unexpectedness measure while keeping the same level of accuracy performance?}

\textbf{RQ3}: \textit{Are the improvements robust to different experimental settings?}

\subsection{Datasets}
We implement our model on three real-world datasets: the Yelp Challenge Dataset Round 12\footnote{https://www.yelp.com/dataset/challenge}, which contains ratings and reviews of users and restaurants; the TripAdvisor Dataset\footnote{http://www.cs.cmu.edu/~jiweil/html/hotel-review.html}, which contains check-in information of users and hotels; and the Video Dataset, which includes the traffic logs we collected from a large-scale industrial video platform. Specifically, we use four days of traffic logs for the training process and the following day for the evaluation process. We list the descriptive statistics of these datasets in Table \ref{statisticalnumber}. To avoid the cold-start and sparsity issues, we filter out users and items that appear less than 5 times in all three datasets.

\begin{table}[h]
\centering
\resizebox{0.6\columnwidth}{!}{
\begin{tabular}{|c|c|c|c|}
\hline
Dataset & \textbf{Yelp} & \textbf{TripAdvisor} & \textbf{Video}\\ \hline
\# of Records & 5,996,996 & 878,561 & 1,155,987 \\ \hline
\# of Items & 188,593 & 576,689 & 287,607 \\ \hline
\# of Users & 1,518,169 & 3,945 & 5,241 \\ \hline
Sparsity & 0.002\% & 0.039\% & 0.077\% \\ \hline
\end{tabular}
}
\caption{Descriptive Statistics of Three Datasets}
\label{statisticalnumber}
\end{table}

\subsection{Parameter Settings}
We perform Bayesian optimization \cite{NIPS2012_4522} to select optimal hyperparameters for the proposed method as well as baseline models. The $\alpha$ is selected as 0.03, where we achieve the optimal balance between the accuracy and unexpectedness measures. In addition, the dimension of the latent embeddings is 128, which is efficient to capture the relations between users and items, as shown in \cite{zhang2017deep}. Detailed parameter settings are further introduced in the next section.

As discussed in Section 4, for three different datasets we select three state-of-the-art embedding approaches accordingly to model the unexpectedness in the latent space. Specifically, the Yelp dataset contains information about explicit users, items and ratings, as well as substantial amounts of meta-information, including text reviews, friendship network, user demographic and geolocation. Thus, it is suitable to be analyzed using Heterogeneous Information Network Embedding (HINE) approach to address the heterogeneous relationships within the Yelp dataset.  Meanwhile, due to the multimodality of video data structure, we utilize the Multimodal Embedding (ME) approach to calculate the unexpectedness between users and videos in the Video dataset. Meanwhile, the TripAdvisor dataset only includes users, items and their associated feature information, which makes the AutoEncoding (AE) approach a reasonable choice for obtaining latent embeddings. 

We point out that, although it could further increase the validity of our approach if we test the same embedding approach on the three datasets, it is not practical to do so. By implementing our model through three different embedding approaches, we illustrate the strength of modeling unexpectedness in the latent space. Note that illustration of this point does not rely on the specific design of embedding approaches. 

\subsection{Training Procedure}
Our proposed latent unexpected recommendation model follows a three-step training procedure: first, we utilize the latent embedding approaches to map users and items into the latent space; then we subsequently calculate the unexpectedness and construct the hybrid utility function for each user; finally, we provide unexpected recommendations based on the hybrid utility function and update our model accordingly.

To obtain the heterogeneous information network embeddings from the Yelp dataset, we extract the users, restaurants and feature labels from the dataset to construct the nodes in the heterogeneous information network. We link the user nodes and items nodes with their associate feature nodes, and we also link the user node with the item node if the user has visited that restaurant before.
We conduct heterogeneous random walk \cite{shi2018heterogeneous} with length 100 starting from each node to generate the sequences of nodes. We repeat this process 10 times. Then we enable skip-gram mechanism following the procedures in \cite{grover2016node2vec} with window size 2, minimal term count 1 and iterations 100 to map the nodes into the latent space, and obtain the corresponding latent embeddings.

To obtain the autoencoder embeddings from the TripAdvisor dataset, we utilize one layer of MLP (Multi-Layer Perceptron) as the encoder to generate latent representations for each user and item, and then use one layer of MLP as decoder to reconstruct the original information. We jointly optimize encoder and decoder to generate the latent embeddings. 

To obtain the multimodal embeddings from the Video dataset, we decompose the input videos into texts, audios and images, where we subsequently apply FCN (Fully-Connected Network) with L2-Norm as regularization term to obtain the latent embeddings separately. Then we concatenate text embeddings, audio embeddings and image embeddings to go through another layer of FCN to generate the final multimodal embeddings.

For performance comparison, we select the deep-learning based Neural Collaborative Filtering (NCF) model \cite{he2017neural} as well as five popular collaborative filtering algorithms including k-Nearest Neighborhood approach (KNN) \cite{altman1992introduction}, Singular Value Decomposition approach (SVD) \cite{sarwar2002incremental}, Co-Clustering approach \cite{george2005scalable}, Non-Negative Matrix Factorization approach (NMF) \cite{lee2001algorithms} and Factorization Machine approach (FM) \cite{rendle2010factorization} to verify robustness of the proposed model. We implement the model in the Python environment using the ''Surprise'',''SciPy'' and ''Gensim'' packages. All experiments are performed on a laptop with 2.50GHz Intel Core i7 and 8GB RAM. We show that the training procedure is time-efficient: it takes 3 hours, 0.5 hours and 1 hours respectively for our proposed model to obtain latent embeddings in the Yelp dataset, the TripAdvisor dataset and the Video dataset. The subsequent unexpected recommendation process takes less than one hour to complete.

\subsection{Evaluation Metrics: Accuracy and Unexpectedness}
To compare the performance of the proposed Latent Closure (LC) method and baseline models, we measure the recommendation results along two dimensions: \textit{accuracy}, in terms of RMSE, MAE, Precision@N and Recall@N metrics \cite{herlocker2004evaluating}, and \textit{unexpectedness}, in terms of Unexpectedness, Serendipity and Diversity metrics \cite{ge2010beyond}. Specifically, we calculate unexpectedness through equation (5) following our proposed definition, while serendipity and diversity are computed following the standard measures in the literature \cite{ziegler2005improving,ge2010beyond}.
\begin{equation}
Serendipity = \frac{RS\& PM \&USEFUL}{RS}
\end{equation}
Serendipity is computed as the percentage of serendipitous recommendations, where \textit{RS} stands for the recommended items using the target model, \textit{PM} stands for the recommendation items using a primitive prediction algorithm (usually selected as the linear regression) and \textit{USEFUL} stands for the items whose utility is above the average level. Diversity is computed as the average intra-list distance.
\begin{equation}
Diversity = \sum_{i \in RS}\sum_{j \neq i \in RS} sim(i,j)
\end{equation}

\subsection{Baseline Models}
We implement several state-of-the-art unexpected recommendation models as baselines and report their performance in terms of aforementioned metrics. The baseline models include SPR, Auralist, DPP, HOM-LIN. Note that we do not include neural network approach because there is no deep-learning based model for unexpected recommendations in the literature.
\begin{itemize}
\item \textbf{SPR \cite{lu2012serendipitous}.} Serendipitous Personalized Ranking is a simple and effective method for serendipitous item recommendation that extends traditional personalized ranking methods by considering item popularity in AUC optimization, which makes the ranking sensitive to the popularity of negative examples.
\item \textbf{Auralist \cite{zhang2012auralist}.} Auralist is a personalized recommendation system that balances between the desired goals of accuracy, diversity, novelty and serendipity simultaneously. Specifically in the music recommendation, the authors combine Artist-based LDA recommendation with two novel components: Listener Diversity and Musical Bubbles. We adjust the algorithm accordingly to fit in our restaurant and hotel recommendation scenario.
\item \textbf{DPP \cite{chen2018fast}} The determinantal point process (DPP) is an elegant probabilistic model of repulsion with applications in various machine learning tasks. The authors propose a fast greedy MAP inference approach for DPP to generate relevant and diverse recommendations.
\item \textbf{HOM-LIN \cite{adamopoulos2015unexpectedness}.} HOM-LIN is the state-of-the-art unexpected recommendation algorithm, where the author propose to define unexpectedness as the distance between items and the expected set of users in the feature space and linearly combine unexpectedness with estimated ratings to provide recommendations. 
\end{itemize}

\subsection{Significant Testing}
To illustrate the differences of recommendation performance between our proposed model and the baseline methods, we conduct significant testing over the experiment results. Specifically, the significance level is determined through rerunning the unexpected recommendation models with random initialization multiple times and conduct Student’s t-test to compute the p-value. We report the significance level together with our results in the next section.

\subsection{Cold-Start Problem}
Note that, the cold start problem is very important in recommender systems. We would like to point out that our proposed unexpected recommender system does not encounter this problem due to the following reasons: First, for the user-side cold start problem, we do not provide unexpected recommendations, as the new users have very few interactions and normally do not face the problem of boredom. Instead, we suggest to provide classical recommendations, which aim at producing similar recommendations to help the users identify and reinforce their interested contents. Second, for the item-side cold start problem, the new item embeddings could be obtained through classical cold start embedding methods \cite{wang2018billion}, and then we could subsequently calculate the unexpectedness and provide unexpected recommendations accordingly.

\section{Results}
In this section, we report the experimental results on three real-world datasets to answer the research questions in Section 5. 

\begin{table}[!]
\begin{tabular}{|c|c|c|c|c|c|c|c|} \hline
\multicolumn{8}{|c|}{Yelp Dataset} \\ \hline
Model & RMSE & MAE & Pre@5 & Rec@5 & Unexp & Ser & Div \\ \hline
NCF+LC & 0.9169 & 0.7078 & \textbf{0.7783*} & \textbf{0.6291*} & 0.1450 & \textbf{0.4905*} & \textbf{0.4178*} \\
FM+LC & 0.9180 & \textbf{0.6888*} & 0.7704 & 0.6278 & 0.1378 & 0.4603 & 0.4164 \\
CC+LC & 0.9514 & 0.7007 & 0.7626 & 0.5926 & 0.1355 & 0.4793 & 0.3961 \\
SVD+LC & 0.9136 & 0.7039 & 0.7722 & 0.6212 & 0.1214 & 0.4630 & 0.3511 \\
NMF+LC & 0.9522 & 0.7026 & 0.7781 & 0.6238 & \textbf{0.1466*} & 0.4894 & 0.4045 \\
KNN+LC & \textbf{0.9133*} & 0.7715 & 0.7674 & 0.6287 & 0.1288 & 0.4380 & 0.3388 \\ \hline
SPR         & 1.0351 & 0.7729 & 0.7692 & 0.6188 & 0.0668 & 0.3720 & 0.2532 \\
Auralist     & 1.0377 & 0.7799 & 0.7678 & 0.6000 & 0.0663 & 0.3637 & 0.2047 \\
HOM-LIN  & 0.9609 & 0.7447 & 0.7621 & 0.6150 & 0.0751 & 0.4329 & 0.3011 \\
DPP          & 1.0288 & 0.7702 & 0.7598 & 0.6012 & 0.0670 & 0.4488 & 0.2488 \\ \hline     
\end{tabular}
\caption{Comparison of unexpected recommendation performance in the Yelp dataset, ''*'' stands for 95\% statistical significance}
\label{result1}

\begin{tabular}{|c|c|c|c|c|c|c|c|c|} \hline
\multicolumn{8}{|c|}{TripAdvisor Dataset} \\ \hline
Model & RMSE & MAE & Pre@5 & Rec@5 & Unexp & Ser & Div \\ \hline
NCF+LC & \textbf{0.9624*} & \textbf{0.7310*} & \textbf{0.7201*} & \textbf{0.9810*} & 0.0586 & 0.4635 & 0.0472 \\
FM+LC & 1.0230 & 0.7450 & 0.7031 & 0.9638 & 0.0581 & \textbf{0.4637*} & 0.0388 \\
CC+LC & 1.0230 & 0.7539 & 0.6887 & 0.9754 & 0.0587 & 0.4629 & \textbf{0.0491*} \\
SVD+LC & 0.9908 & 0.7519 & 0.7093 & 0.9569 & 0.0585 & 0.4614 & 0.0477 \\
NMF+LC & 1.0280 & 0.7594 & 0.6864 & 0.9735 & 0.0584 & 0.4629 & 0.0488 \\
KNN+LC & 0.9981 & 0.7493 & 0.6909 & 0.9743 & \textbf{0.0588*} & 0.4625 & 0.0488 \\ \hline
SPR          & 1.0328 & 0.8008 & 0.6395 & 0.9325 & 0.0474 & 0.3593 & 0.0375 \\
Auralist      & 1.0318 & 0.7997 & 0.6460 & 0.9390 & 0.0473 & 0.3462 & 0.0355 \\
HOM-LIN   & 1.0298 & 0.7902 & 0.6420 & 0.9418 & 0.0572 & 0.3729 & 0.0411 \\
DPP           & 1.0304 & 0.8158 & 0.6264 & 0.9303 & 0.0464 & 0.3245 & 0.0311 \\ \hline
\end{tabular}
\caption{Comparison of unexpected recommendation performance in the TripAdvisor dataset, ''*'' stands for 95\% statistical significance}
\label{result2}

\begin{tabular}{|c|c|c|c|c|c|c|c|c|} \hline
\multicolumn{8}{|c|}{Video Dataset} \\ \hline
Model & RMSE & MAE & Pre@5 & Rec@5 & Unexp & Ser & Div \\ \hline
NCF+LC & \textbf{0.3810*} & \textbf{0.2854*} & 0.2560 & 0.3615 & 0.7070 & 0.9830 & 0.2538 \\
FM+LC & 0.3924 & 0.3044 & 0.2498 & 0.3265 & \textbf{0.7096*} & \textbf{0.9833*} & 0.2510 \\
CC+LC & 0.4167 & 0.3296 & 0.2569 & \textbf{0.3676*} & 0.7053 & 0.9815 & 0.2519 \\
SVD+LC & 0.3888 & 0.2862 & 0.2455 & 0.3253 & 0.7018 & 0.9810 & 0.2412 \\
NMF+LC & 0.4405 & 0.3330 & 0.2494 & 0.3439 & 0.6999 & 0.9792 & 0.2450 \\
KNN+LC & 0.4088 & 0.3091 & \textbf{0.2608*} & 0.3212 & 0.9814 & 0.9801 & \textbf{0.2558*} \\ \hline
SPR     & 0.4610 & 0.3638 & 0.2298 & 0.2870 & 0.6300 & 0.9593 & 0.2137 \\
Auralist & 0.4515 & 0.3610 & 0.2304 & 0.2890 & 0.6462 & 0.9462 & 0.1980 \\
HOM-LIN  & 0.4498 & 0.3608 & 0.2310 & 0.2912 & 0.6732 & 0.9473 & 0.2154 \\
DPP      & 0.4770 & 0.3670 & 0.2271 & 0.2870 & 0.6593 & 0.9328 & 0.2154 \\ \hline
\end{tabular}

\caption{Comparison of unexpected recommendation performance in the Video dataset, ''*'' stands for 95\% statistical significance}
\label{result3}
\end{table}

\subsection{Unexpected Recommendation Performance}
To start with, we compare the recommendation performance of proposed latent unexpectedness with baseline unexpected recommendation models. Specifically, the proposed LC method provides unexpected recommendations through $Utility_{u,i}= EstRating_{u,i} + \alpha *Unexp_{u,i}$ where unexpectedness is calculate using Latent HyperSphere introduced in Section 3.2 and estimated ratings are computed through deep-learning based method Neural Collaborative Filtering (NCF) and five other popular collaborative filtering algorithms Factorization Machine (FM), CoClustering (CC), Singular Value Decomposition (SVD), Non-negative Matrix Factorization (NMF) and K-Nearest Neighbor (KNN). We denote the corresponding unexpected recommendations provided through hybrid utility functions as NCF+LC, FM+LC, CC+LC, SVD+LC, NMF+LC and KNN+LC accordingly.

As shown in Table \ref{result1}, \ref{result2} and \ref{result3}, by utilizing the proposed latent modeling of unexpectedness, all six unexpected recommendation models consistently and significantly outperforms the baseline methods in both accuracy and unexpectedness measures. Specifically, we observe an average increase of 5.21\% in RMSE, 8.11\% in MAE, 1.14\% in Precision, 1.57\% in Recall, 48.77\% in Unexpectedness, 8.30\% in Serendipity and 27.69\% in Diversity compared to the second best baseline model in the Yelp dataset. That is to say, the proposed latent modeling of unexpectedness enables us to provide more unexpected and more useful recommendations at the same time. Also, we show that the superiority of latent unexpectedness is robust to the specific selection of collaborative filtering algorithms, as we obtain significant increase of performance measures in all six algorithms and do not observe any significant difference in unexpectedness metric within these methods.

\begin{table}[!]
\begin{tabular}{|c|c|c|c|c|c|c|c|} \hline
Model & RMSE & MAE & Pre@5 & Rec@5 & Unexp & Ser & Div \\ \hline
NCF & 0.9154 & 0.7070 & 0.7761 & 0.6318 & 0.0492 & 0.1666 & 0.3492 \\
NCF+LC & 0.9169 & 0.7078 & 0.7783 & 0.6291 & \textbf{0.1450*} & \textbf{0.4905*} & \textbf{0.4178*} \\ \hline
FM & 0.9197 & 0.6815 & 0.7699 & 0.6223 & 0.0326 & 0.0978 & 0.0135 \\
FM+LC & 0.9180 & 0.6888 & 0.7704 & 0.6278 & \textbf{0.1378*} & \textbf{0.4603*} & \textbf{0.4164*} \\ \hline
CC & 0.9499 & 0.7040 & 0.7655 & 0.5913 & 0.0338 & 0.1595 & 0.3106 \\
CC+LC & 0.9514 & 0.7007 & 0.7626 & 0.5926 & \textbf{0.1355*} & \textbf{0.4793*} & \textbf{0.3961*} \\ \hline
SVD & 0.9132 & 0.7071 & 0.7792 & 0.6244 & 0.0457 & 0.1352 & 0.0479 \\
SVD+LC & 0.9136 & 0.7039 & 0.7722 & 0.6212 & \textbf{0.1214*} & \textbf{0.4630*} & \textbf{0.3511*} \\ \hline
NMF & 0.9533 & 0.7081 & 0.7797 & 0.6318 & 0.0333 & 0.1954 & 0.3268 \\
NMF+LC & 0.9522 & 0.7026 & 0.7781 & 0.6238 &\textbf{0.1466*} & \textbf{0.4894*} & \textbf{0.4045*} \\ \hline
KNN & 0.9123 & 0.7748 & 0.7687 & 0.6285 & 0.0448 & 0.0977 & 0.0129 \\
KNN+LC & 0.9133 & 0.7715 & 0.7674 & 0.6287 & \textbf{0.1288*} & \textbf{0.4380*} & \textbf{0.3388*} \\ \hline 
\end{tabular}
\bigskip
\subfloat[RMSE]{\includegraphics[width=.25\textwidth]{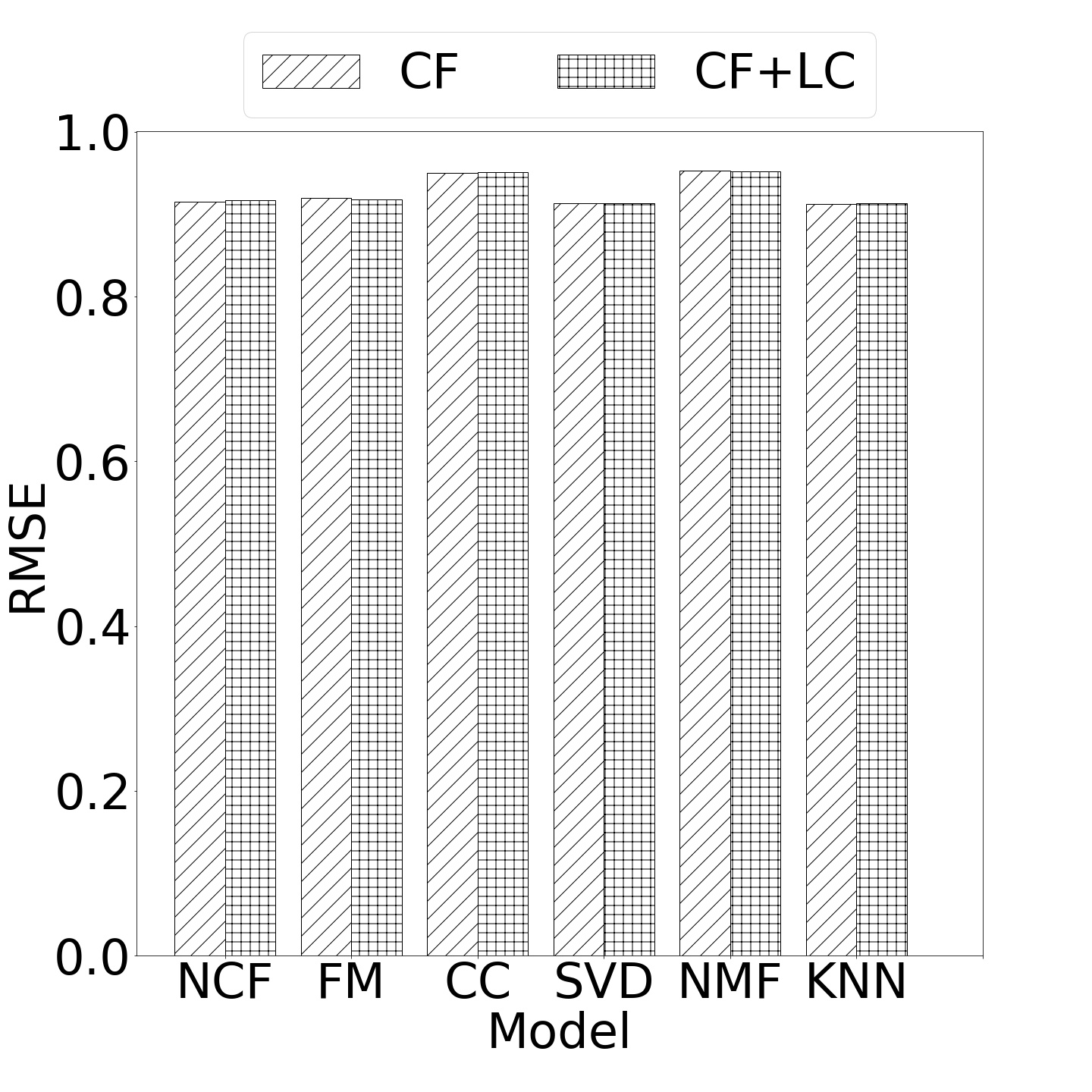}\label{fig:rmse}}
\subfloat[MAE]{\includegraphics[width=.25\textwidth]{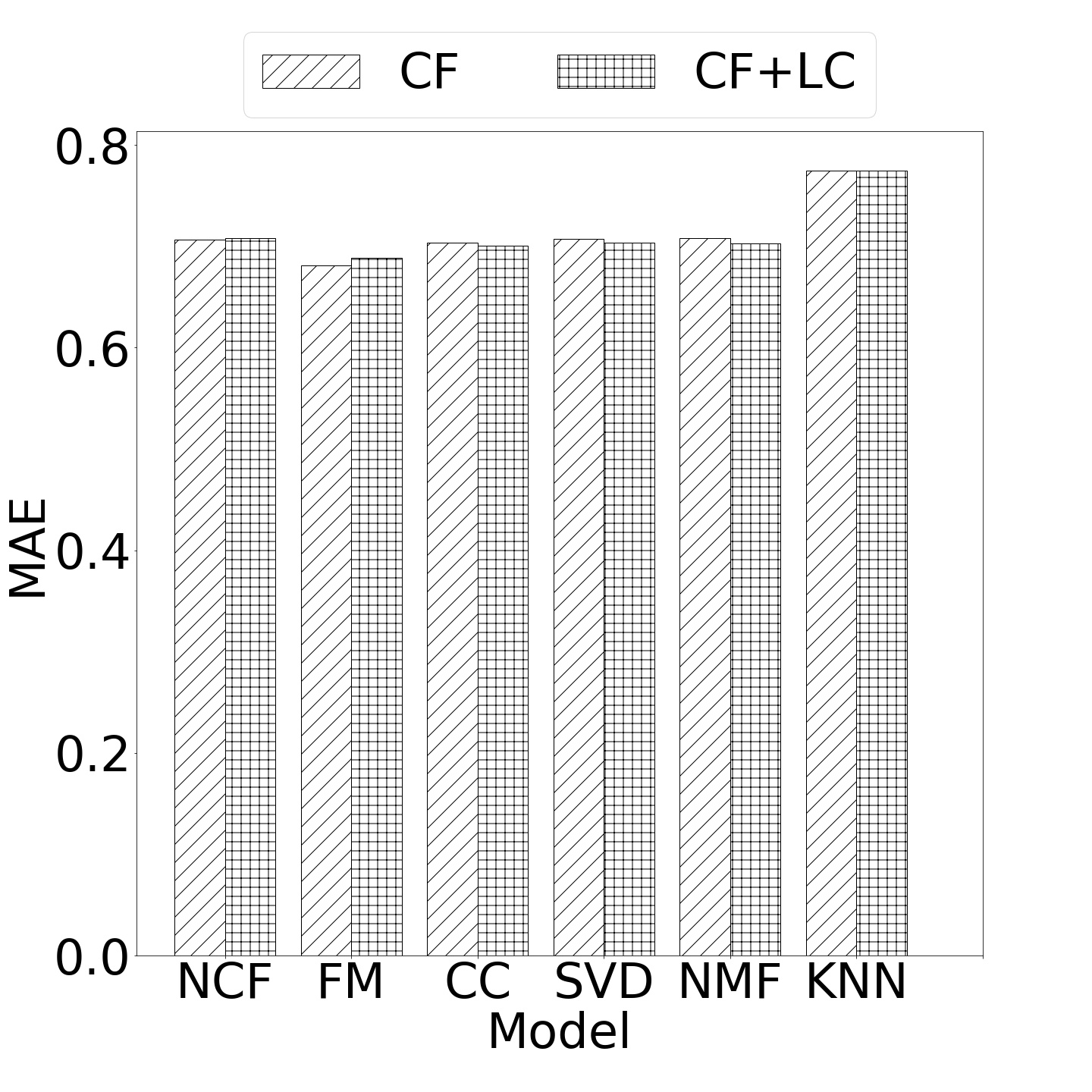}\label{fig:mae}}
\subfloat[Unexpectedness]{\includegraphics[width=.25\textwidth]{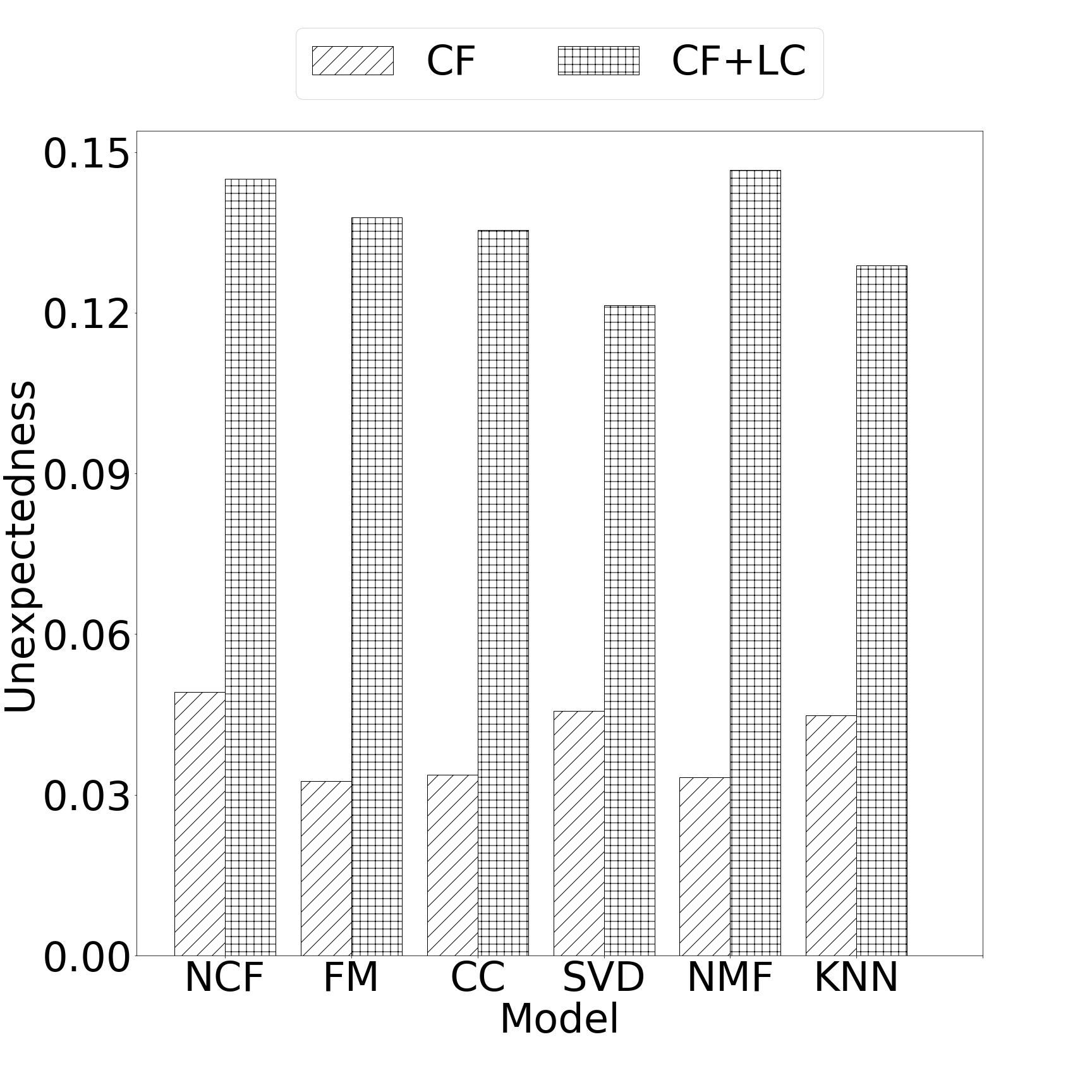}\label{fig:unexp}}
\subfloat[Serendipity]{\includegraphics[width=.25\textwidth]{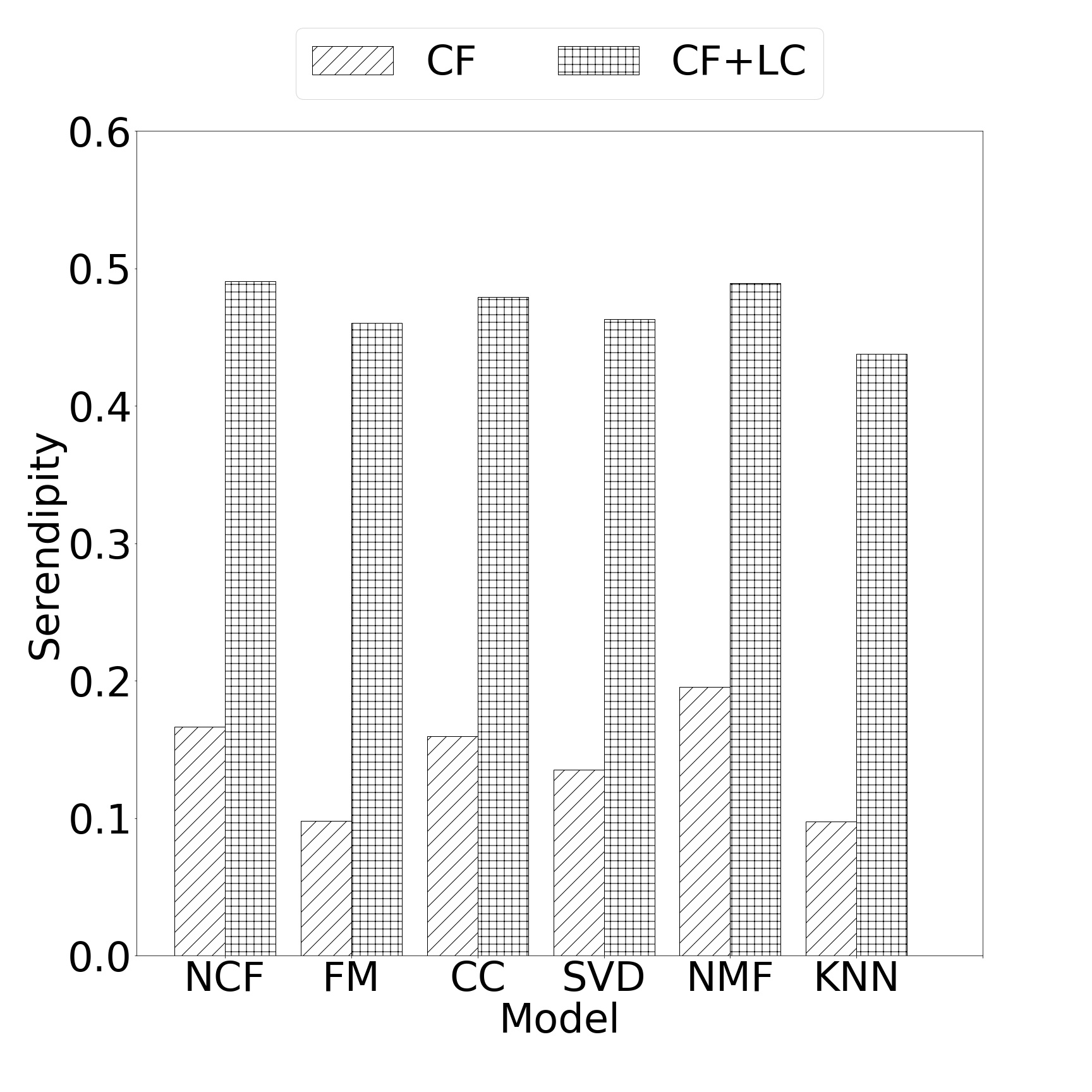}\label{fig:seren}}
\caption{Comparison of recommendation performance with and without unexpectedness in the Yelp dataset, ''*'' stands for 95\% statistical significance; we observe significant improvements in unexpectedness measures in (c) and (d), while no significant change in accuracy measures in (a) and (b) at the same time.}
\label{tradeoff1}
\end{table}

\begin{table}[t]
\begin{tabular}{|c|c|c|c|c|c|c|c|c|} \hline
Model & RMSE & MAE & Pre@5 & Rec@5 & Unexp & Ser & Div \\ \hline
NCF & 0.9588 & 0.7291 & 0.7230 & 0.9815 & 0.0222 & 0.3960 & 0.0010 \\
NCF+LC & 0.9624 & 0.7310 & 0.7201 & 0.9810 & \textbf{0.0586*} & \textbf{0.4635*} & \textbf{0.0472*} \\ \hline
FM & 1.0105 & 0.7440 & 0.7068 & 0.9590 & 0.0222 & 0.3979 & 0.0017 \\
FM+LC & 1.0230 & 0.7450 & 0.7031 & 0.9638 & \textbf{0.0581*} & \textbf{0.4637*} & \textbf{0.0388*} \\ \hline
CC & 1.0178 & 0.7543 & 0.6845 & 0.9732 & 0.0234 & 0.3973 & 0.0015 \\
CC+LC & 1.0230 & 0.7539 & 0.6887 & 0.9754 & \textbf{0.0587*} & \textbf{0.4629*} & \textbf{0.0491*} \\ \hline
SVD & 0.9868 & 0.7533 & 0.7010 & 0.9565 & 0.0231 & 0.3967 & 0.0006 \\
SVD+LC & 0.9908 & 0.7519 & 0.7093 & 0.9569 & \textbf{0.0585*} & \textbf{0.4614*} & \textbf{0.0477*} \\ \hline
NMF & 1.0241 & 0.7609 & 0.6850 & 0.9681 & 0.0227 & 0.3979 & 0.0010 \\
NMF+LC & 1.0280 & 0.7594 & 0.6864 & 0.9735 & \textbf{0.0584*} & \textbf{0.4629*} & \textbf{0.0488*} \\ \hline
KNN & 0.9940 & 0.7531 & 0.6969 & 0.9689 & 0.0233 & 0.3979 & 0.0019 \\
KNN+LC & 0.9981 & 0.7493 & 0.6909 & 0.9743 & \textbf{0.0588*} & \textbf{0.4625*} & \textbf{0.0488*} \\ \hline
\end{tabular}
\bigskip
\subfloat[RMSE]{\includegraphics[width=.25\textwidth]{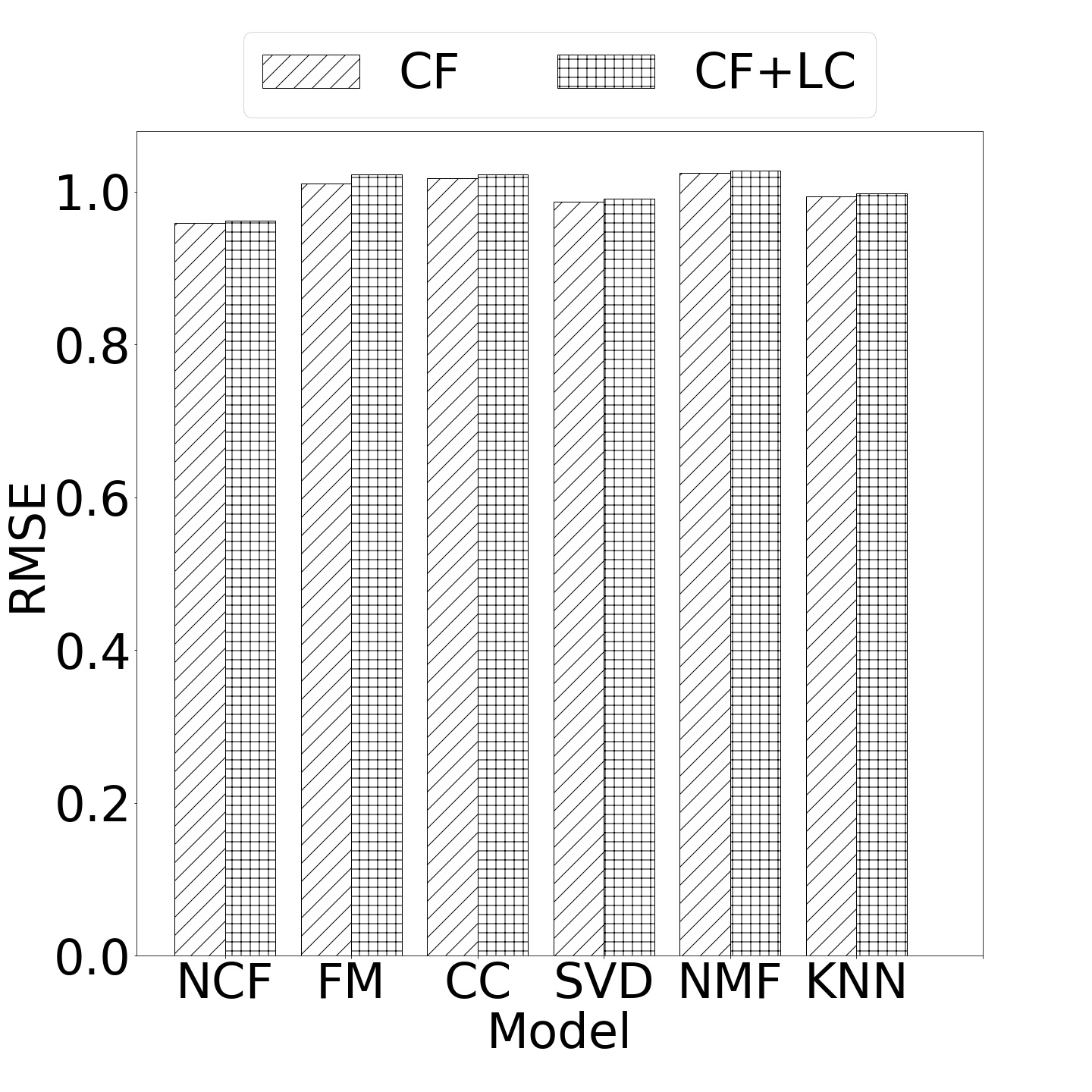}\label{fig:rmse}}
\subfloat[MAE]{\includegraphics[width=.25\textwidth]{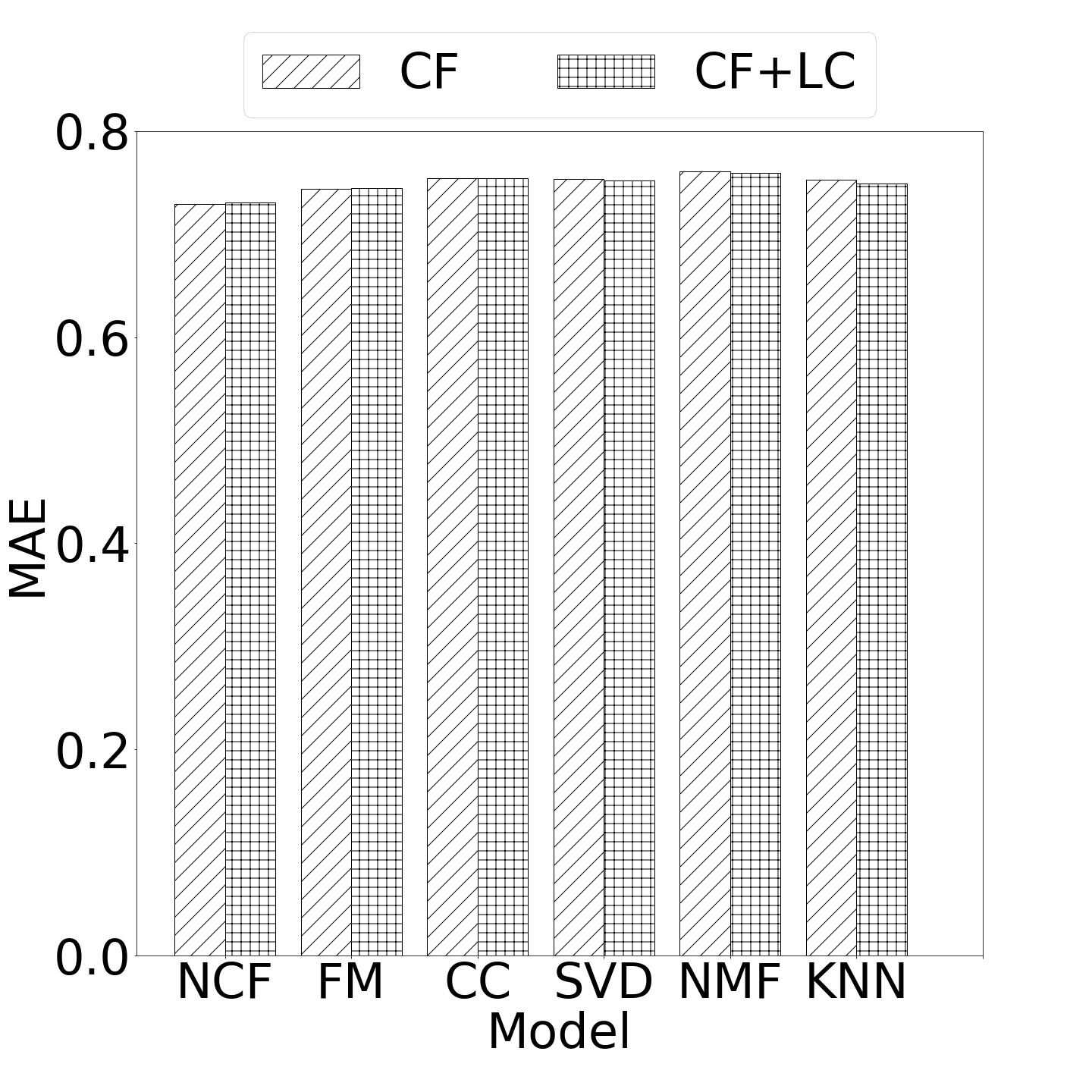}\label{fig:mae}}
\subfloat[Unexpectedness]{\includegraphics[width=.25\textwidth]{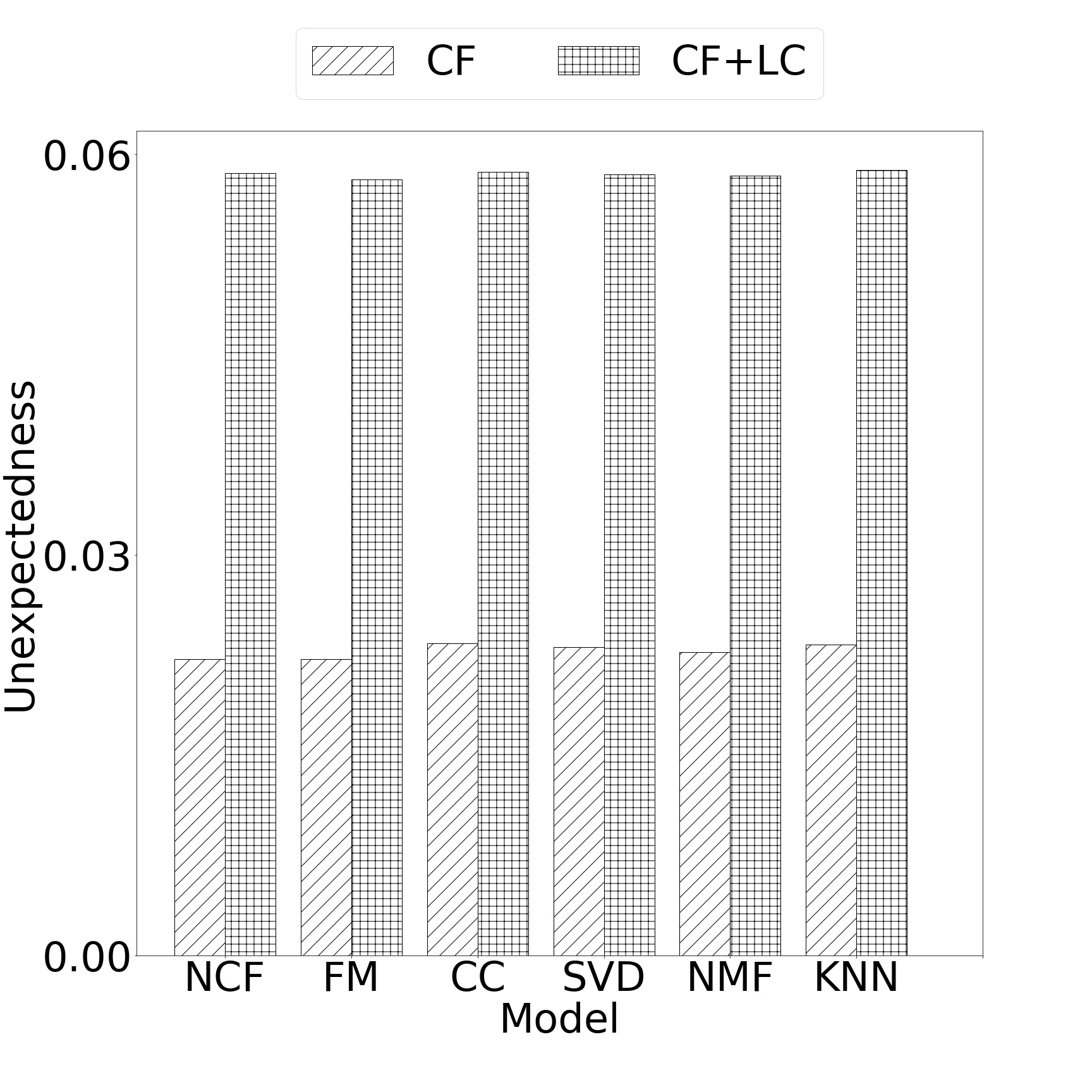}\label{fig:unexp}}
\subfloat[Serendipity]{\includegraphics[width=.25\textwidth]{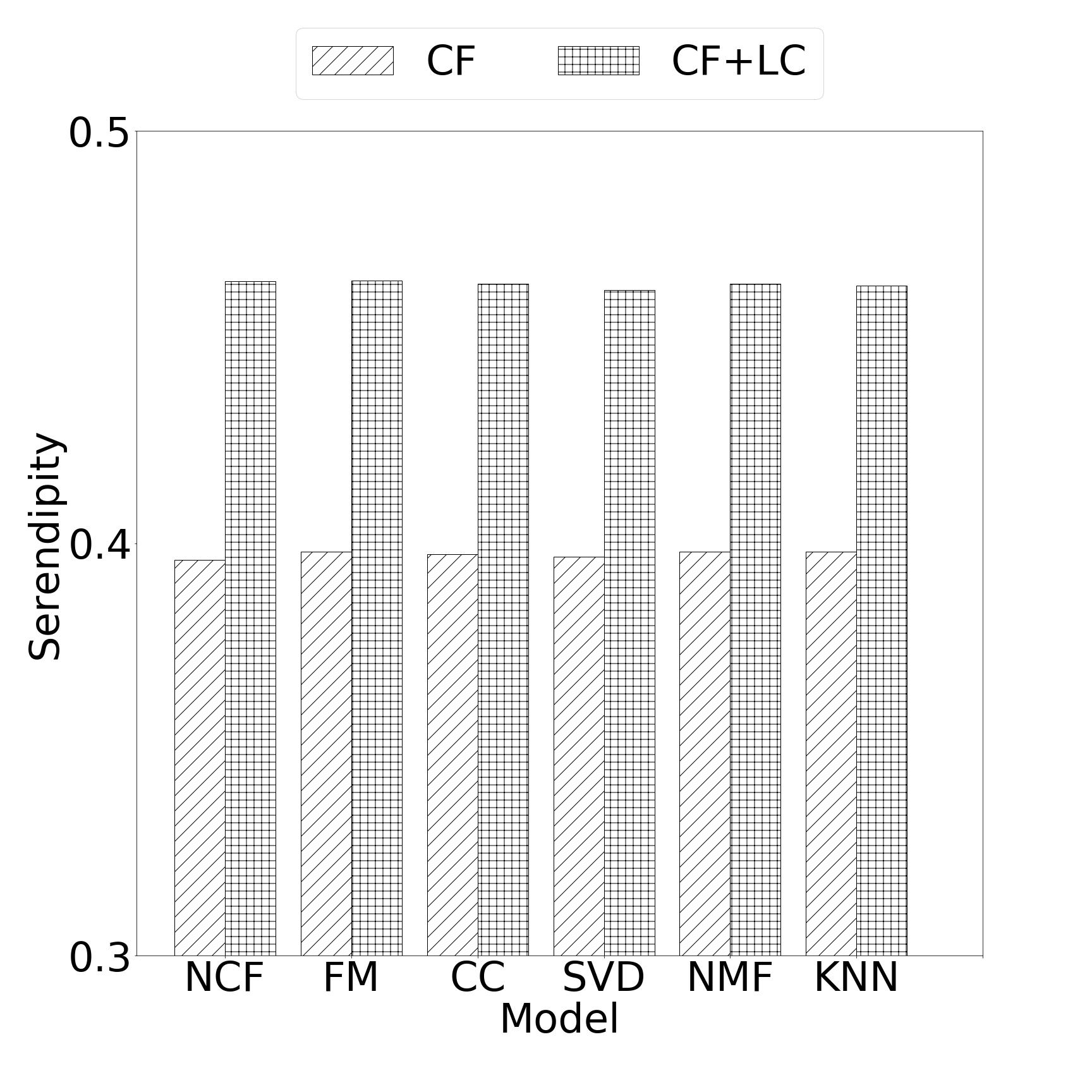}\label{fig:seren}}
\caption{Comparison of recommendation performance with and without unexpectedness in the TripAdvisor dataset, ''*'' stands for 95\% statistical significance; we observe significant improvements in unexpectedness measures in (c) and (d), while no significant change in accuracy measures in (a) and (b) at the same time.}
\label{tradeoff2}
\end{table}

\begin{table}[t]
\begin{tabular}{|c|c|c|c|c|c|c|c|c|} \hline
Model & RMSE & MAE & Pre@5 & Rec@5 & Unexp & Ser & Div \\ \hline
NCF & 0.3815 & 0.2877 & 0.2544 & 0.3630 & 0.6402 & 0.8678 & 0.2317 \\
NCF+LC & 0.3810 & 0.2854 & 0.2560 & 0.3615 & \textbf{0.7070*} & \textbf{0.9830*} & \textbf{0.2538*} \\ \hline
FM & 0.3920 & 0.3013 & 0.2472 & 0.3280 & 0.6398 & 0.8552 & 0.2396 \\
FM+LC & 0.3924 & 0.3044 & 0.2498 & 0.3265 & \textbf{0.7096*} & \textbf{0.9833*} & \textbf{0.2510*} \\ \hline
CC & 0.4129 & 0.3302 & 0.2560 & 0.3646 & 0.6479 & 0.8382 & 0.2362 \\
CC+LC & 0.4167 & 0.3296 & 0.2569 & 0.3676 & \textbf{0.7053*} & \textbf{0.9815*} & \textbf{0.2519*} \\ \hline
SVD & 0.3806 & 0.2895 & 0.2392 & 0.3232 & 0.6495 & 0.8480 & 0.2346 \\
SVD+LC & 0.3888 & 0.2862 & 0.2455 & 0.3253 & \textbf{0.7018*} & \textbf{0.9810*} & \textbf{0.2412*} \\ \hline
NMF & 0.4462 & 0.3285 & 0.2480 & 0.3391 & 0.6548 & 0.8655 & 0.2385 \\
NMF+LC & 0.4405 & 0.3330 & 0.2494 & 0.3439 & \textbf{0.6999} & \textbf{0.9792*} & \textbf{0.2450*} \\ \hline
KNN & 0.4103 & 0.3048 & 0.2531 & 0.3173 & 0.6416 & 0.8632 & 0.2385 \\
KNN+LC & 0.4088 & 0.3091 & 0.2608 & 0.3212 & \textbf{0.7014*} & \textbf{0.9814*} & \textbf{0.2558*} \\ \hline
\end{tabular}
\bigskip
\subfloat[RMSE]{\includegraphics[width=.25\textwidth]{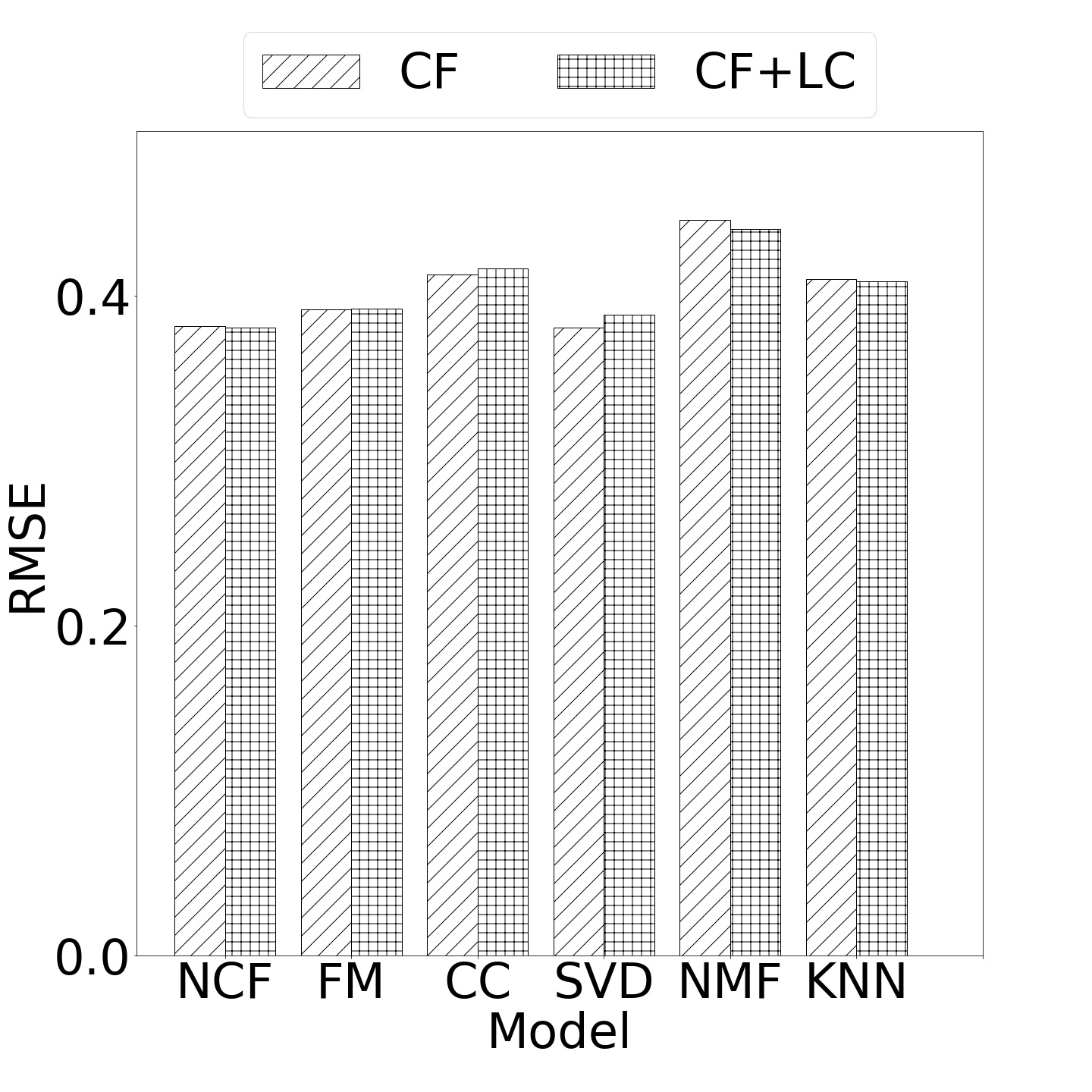}\label{fig:rmse}}
\subfloat[MAE]{\includegraphics[width=.25\textwidth]{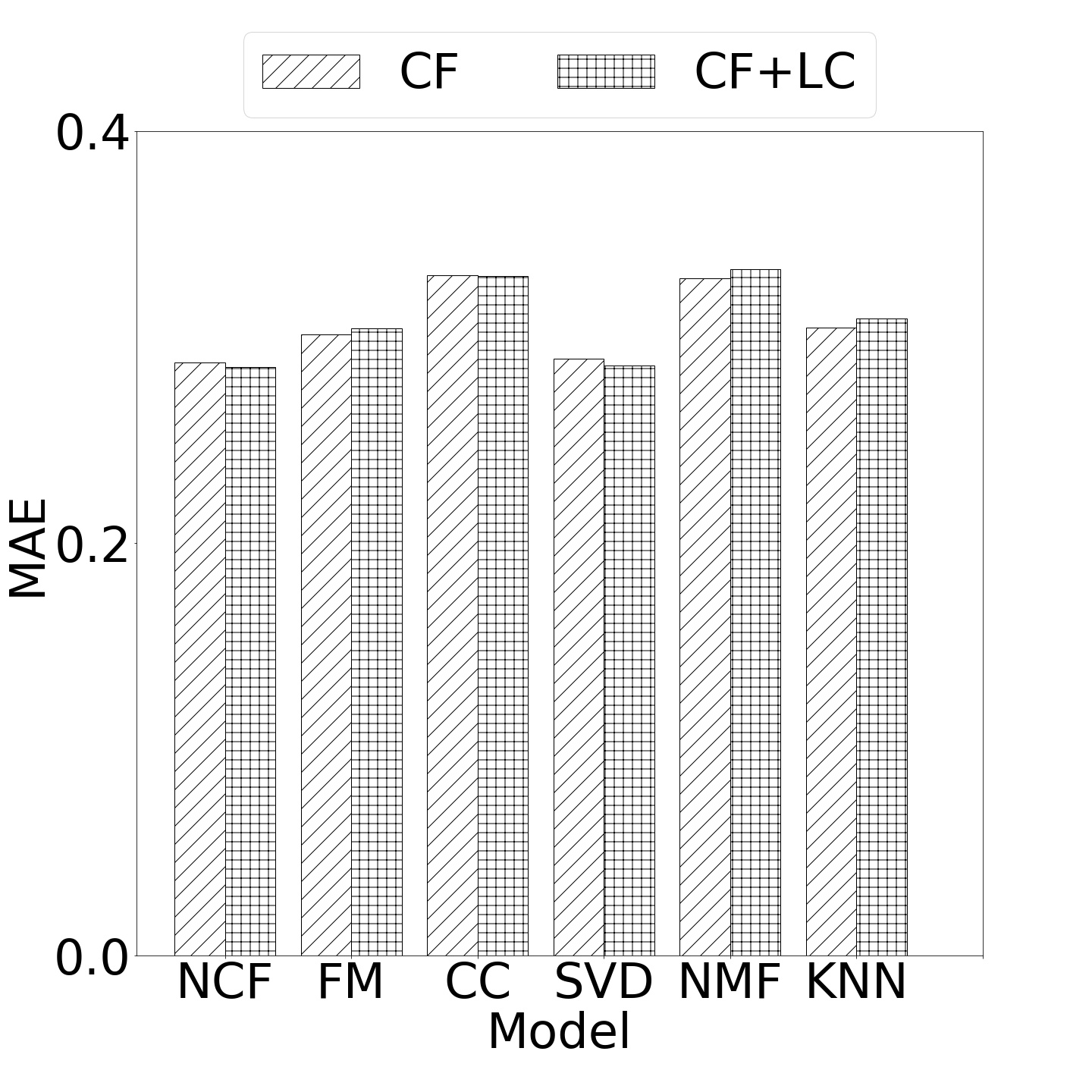}\label{fig:mae}}
\subfloat[Unexpectedness]{\includegraphics[width=.25\textwidth]{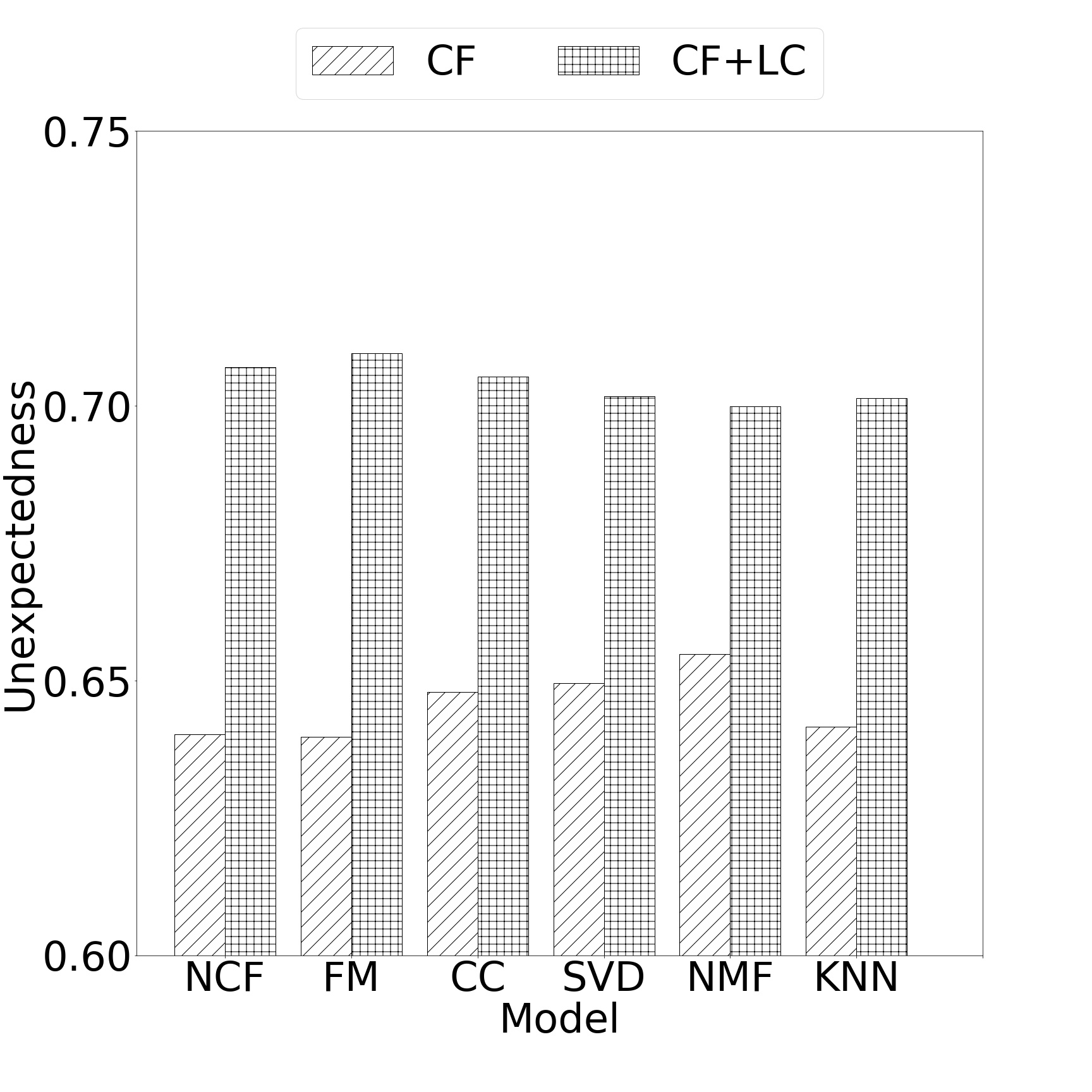}\label{fig:unexp}}
\subfloat[Serendipity]{\includegraphics[width=.25\textwidth]{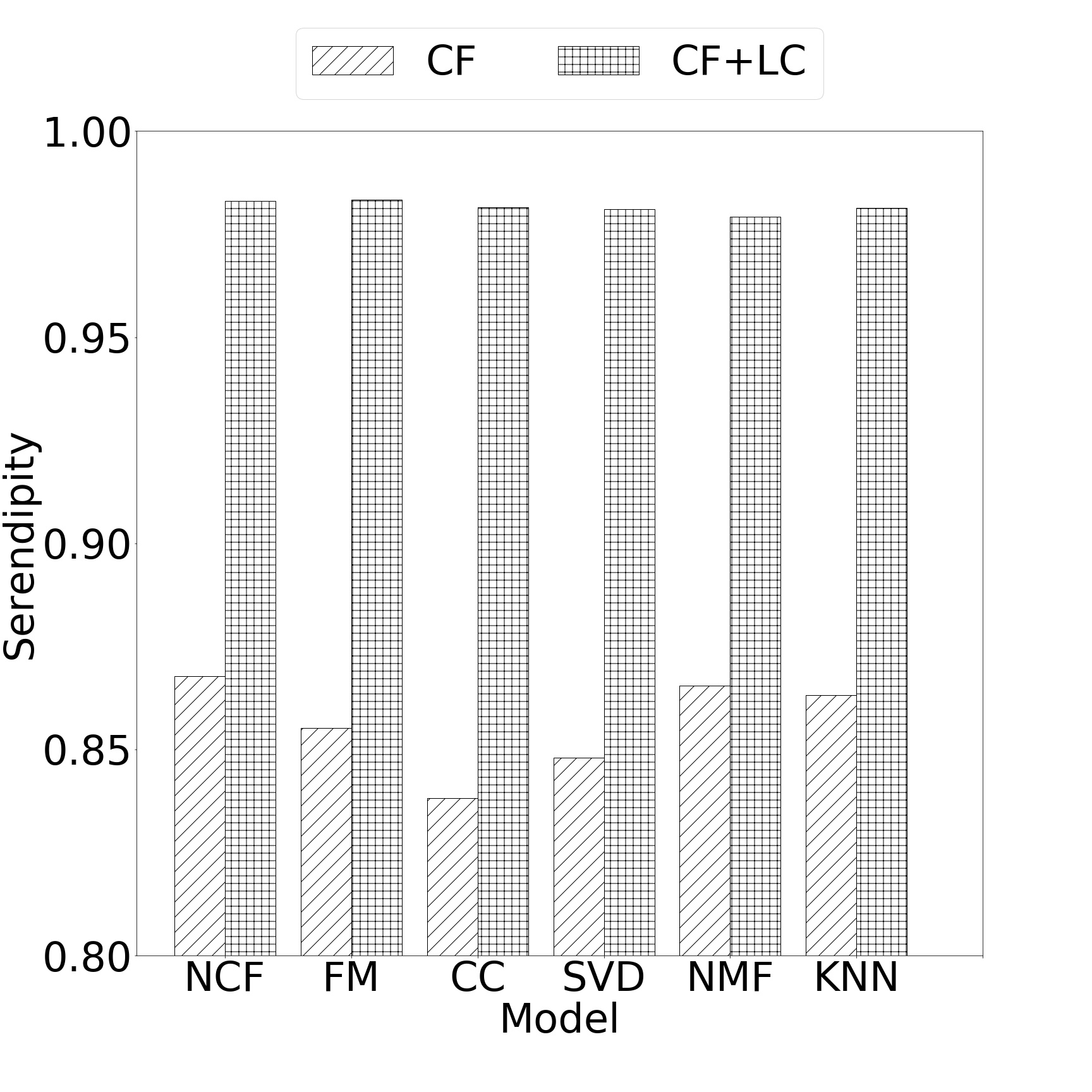}\label{fig:seren}}
\caption{Comparison of recommendation performance with and without unexpectedness in the Video dataset, ''*'' stands for 95\% statistical significance; we observe significant improvements in unexpectedness measures in (c) and (d), while no significant change in accuracy measures in (a) and (b) at the same time.}
\label{tradeoff3}
\end{table}

\subsection{Improving Unexpectedness while Keeping Accuracy}
As we discuss in the previous section, an important problem with incorporating unexpectedness into recommendations is the trade-off between accuracy and novelty measures \cite{zolaktaf2018generic,zhou2010solving}, which is crucial to the practical use of unexpected recommendations. In this section, we compare the unexpected recommendation performance using hybrid utility functions with those classical recommender systems that provide recommendations based on estimated ratings only.

As shown in Table \ref{tradeoff1}, \ref{tradeoff2}, \ref{tradeoff3} and the corresponding plots, when including unexpectedness in the recommendation process, we consistently obtain significant improvements in terms of unexpectedness, serendipity and diversity measures, while we do not witness any loss in the accuracy measures. Therefore, we show that it is indeed the proposed latent closure approach that enables us to provide useful and unexpected recommendations simultaneously. It is crucial for the successful deployment of unexpected recommendation models in the industrial applications.

In addition, we study the impact of the hyperparameter $\alpha$ in Equation (5) that controls for the degree of unexpectedness and usefulness in the hybrid utility function. Typically a higher value of $\alpha$ indicates that the recommendation model is in favor of unexpected recommendations over useful recommendations, while a lower value of $\alpha$ tends to recommend more useful items as opposed to unexpected items. 

\begin{figure*}[h]
\centering
\subfloat[Yelp]{\includegraphics[width=.33\textwidth]{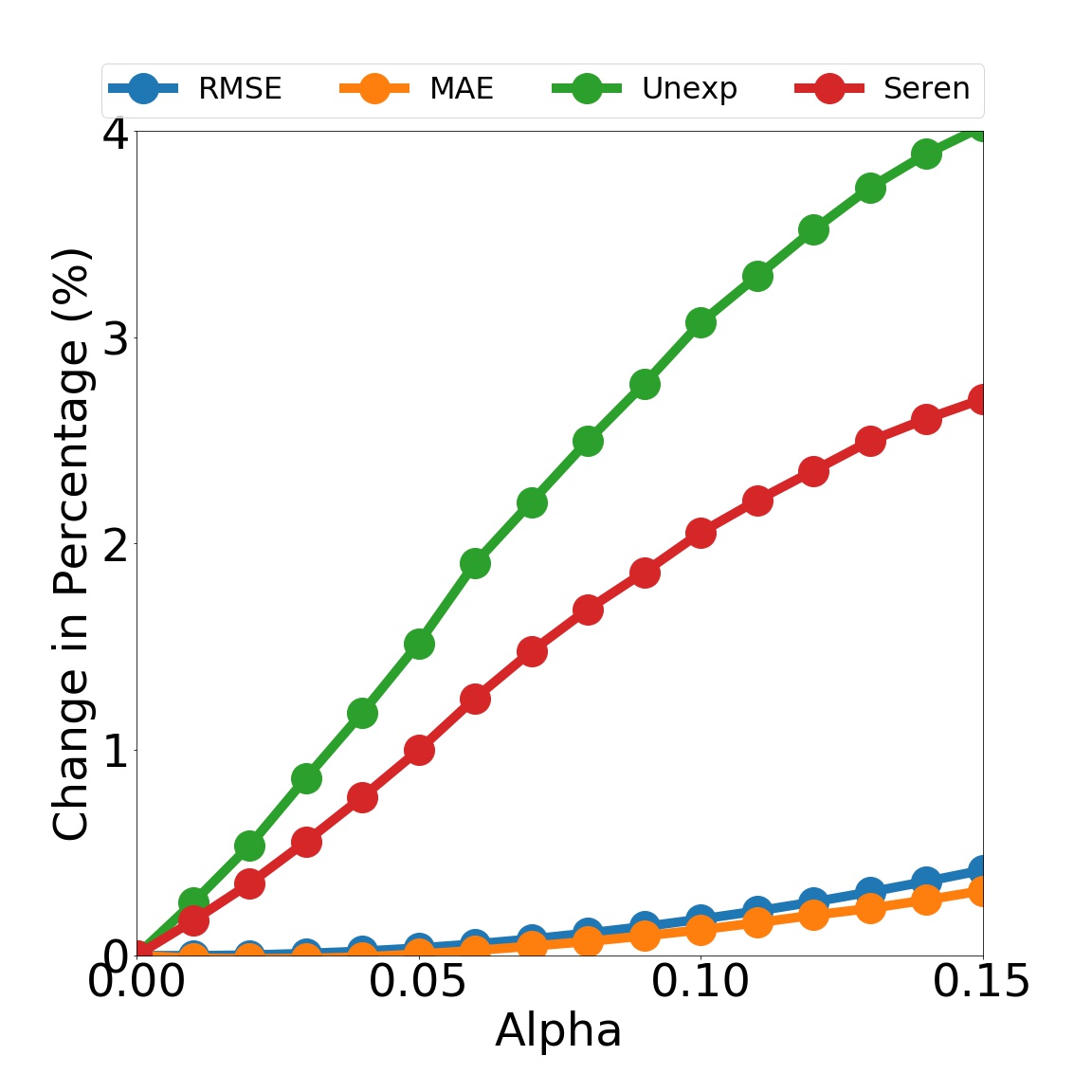}\label{fig:yelp}}
\subfloat[TripAdvisor]{\includegraphics[width=.33\textwidth]{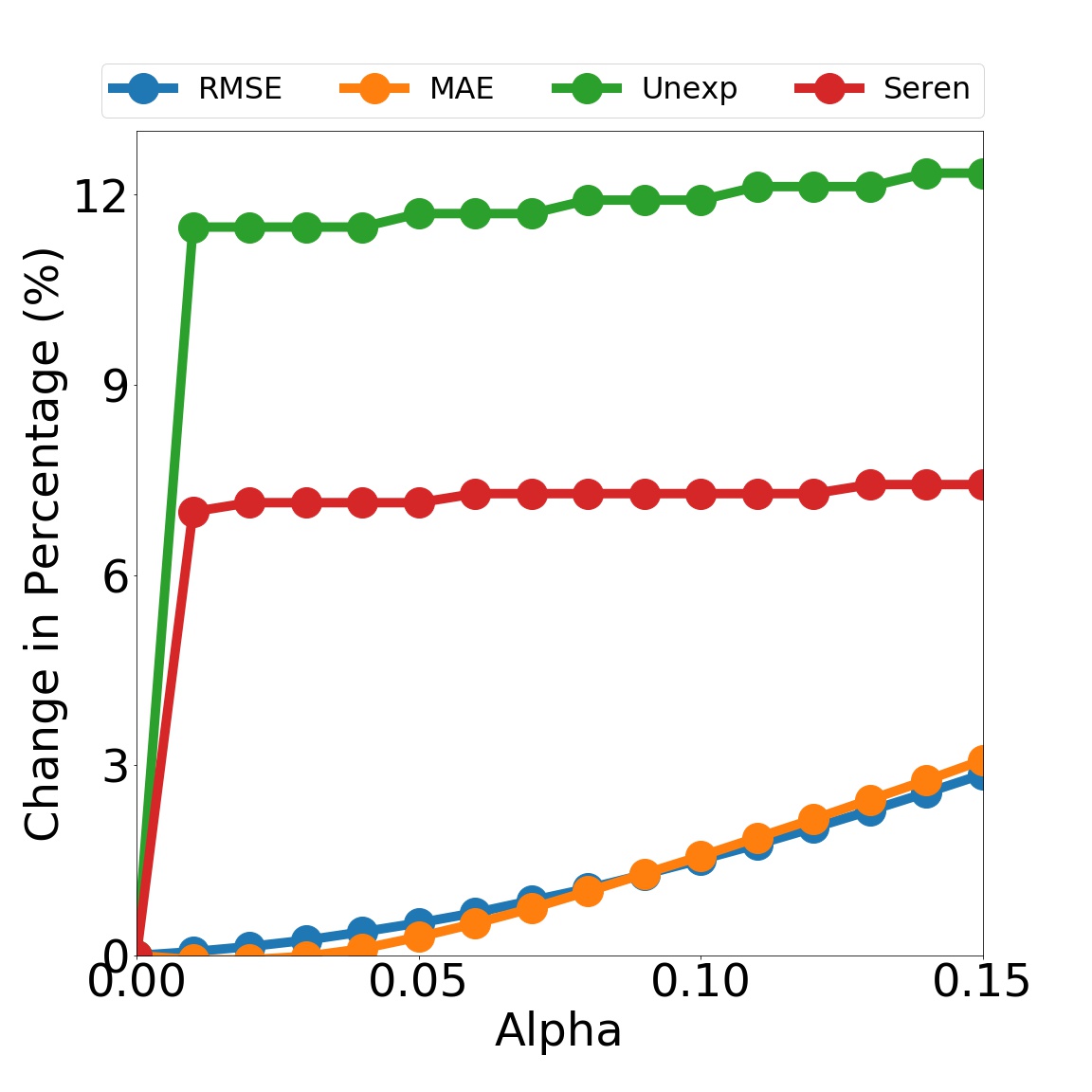}\label{fig:tripadvisor}}
\subfloat[Video]{\includegraphics[width=.33\textwidth]{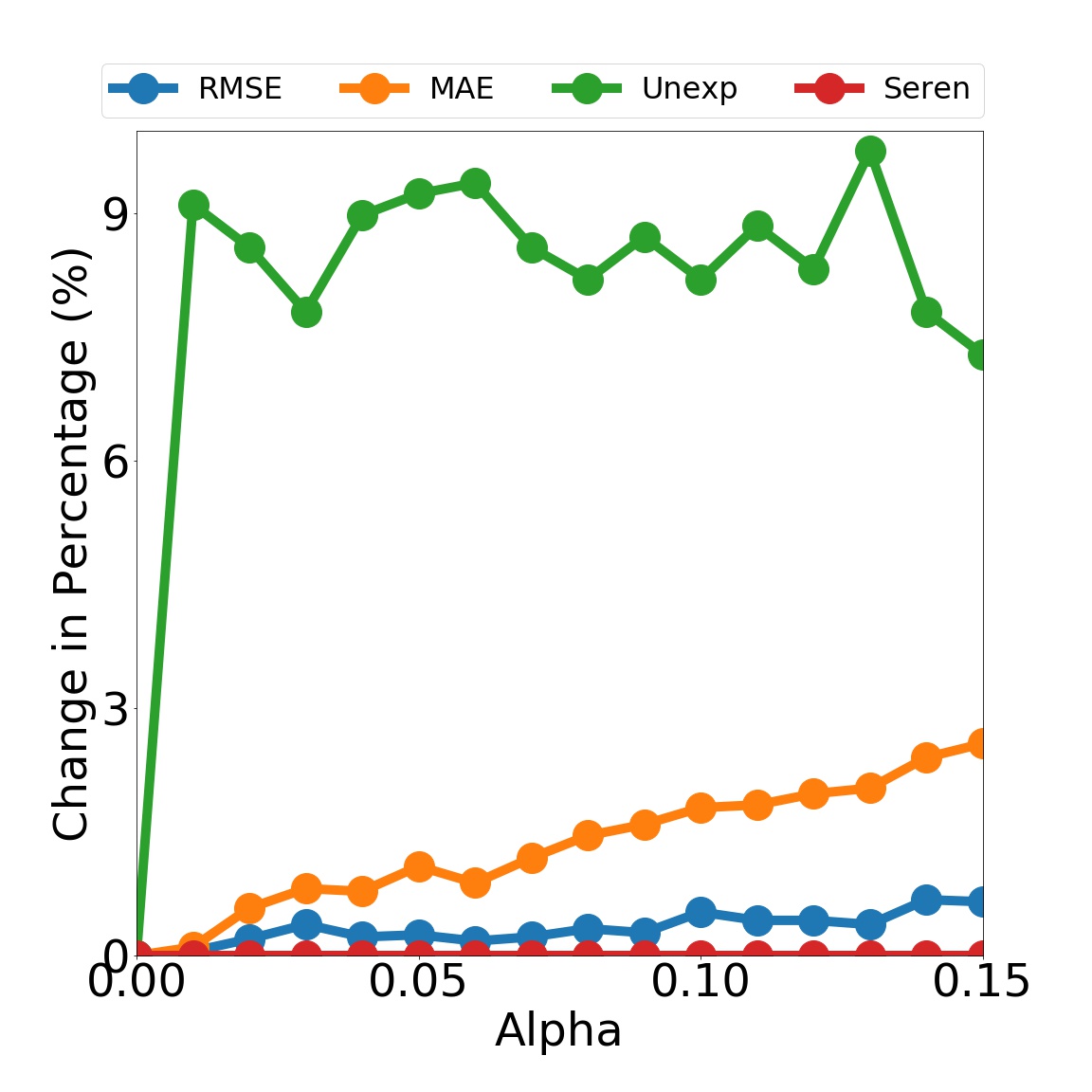}\label{fig:industrial}}
\caption{Comparison of Accuracy-Novelty Trade-off}
\label{tradeoff}
\end{figure*}

We plot the change of accuracy and novelty measures with respect to different $\alpha$ value in Figure \ref{tradeoff}. This figure illustrates that when we select relatively small value of $\alpha$, (e.g., $\alpha$=0.03) we can obtain significant amount of increase in unexpectedness (8.40\%, 10.81\% and 9.10\% respectively in three datasets) while the decrease of accuracy performance is not statistically significant for all three datasets. It is also worth noticing that if we select a large value of $\alpha$, we might risk deteriorating the accuracy performance of recommendations significantly.

\subsection{Robustness Analysis}
In this paper, we show that the proposed latent modeling of unexpectedness significantly improves recommendation performance and provide indeed unexpected recommendations. In total, we conduct the experiments on 3 different datasets. using 3 different latent embedding approaches, 6 different collaborative filtering algorithms, 7 different evaluation metrics and 3 different geometric structures for modeling unexpectedness, resulting in 378 experimental settings, where all 378 results are in supportive of our claims. We observe significant improvements in unexpectedness, serendipity and diversity measures, while we do not witness any loss in accuracy measures compared to plain collaborative filtering algorithms that do not include unexpectedness during the recommendation process. In addition, when compared to baseline unexpected recommendation models, our model significantly outperforms them in both accuracy and unexpectedness measures.

To sum up, the superiority of latent modeling of unexpectedness is robust to
\begin{itemize}
\item \textbf{Various Datasets} We conduct the experiments in three different datasets: Yelp dataset, TripAdvisor dataset and Video dataset, where we obtain consistent improvements in all three datasets.
\item \textbf{Multiple Latent Embedding Approaches} To construct the unexpectedness in the latent space, we utilize three state-of-the-art latent embedding approaches: Heterogeneous Information Network Embeddings (HINE), Autoencoder Embeddings (AE) and Multimodal Embeddings (ME) and obtain similar superior recommendation performance over baseline models.
\item \textbf{Specific Collaborative Filtering Algorithms} We select six representative collaborative filtering algorithms to estimate user ratings and form the hybrid utility function accordingly. These methods include the deep-learning based approach NCF and five other popular models FM, CC, SVD, NMF and KNN. The latent modeling of unexpectedness enables each collaborative filtering algorithm to provide more unexpected recommendations without losing any accuracy measure.
\item \textbf{Selective Evaluation Metrics} We evaluate the recommendation performance using accuracy measures RMSE, MAE, Precision, Recall and unexpectedness measures Unexpectedness, Serendipity, Diversity. The proposed model significantly outperforms baseline unexpected recommendation models in all these seven metrics.
\item \textbf{Different Geometric Shapes of Latent Closures} As discussed in Section 3.2,  there are three common geometric structures in high-dimensional latent space that are suitable for modeling the closure of latent embeddings: Latent HyperSphere (LHS), Latent HyperCube (LHC) and Latent Convex Hull (LCH). We calculate unexpectedness using the three structures separately and provide unexpected recommendations accordingly. As shown in Table \ref{structure1}, \ref{structure2} and \ref{structure3}, the specific selection of geometric structure does not influence the recommendation performance, as we get similar results and neither approach dominates the other two. Instead, it is really the latent modeling of unexpectedness that contributes to the significant improvements of recommendation performance.
\end{itemize}

\begin{table}
\begin{tabular}{|c|c|c|c|c|c|c|c|} \hline
Model & RMSE & MAE & Pre@5 & Rec@5 & Unexp & Ser & Div \\ \hline
NCF+LCH & 0.9158 & 0.7076 & 0.7798 & 0.6308 & 0.1478 & 0.4889 & 0.4170 \\
NCF+LHS & 0.9169 & 0.7078 & 0.7783 & 0.6291 & 0.1450 & 0.4905 & 0.4178 \\
NCF+LHC & 0.9180 & 0.7013 & 0.7725 & 0.6270 & 0.1478 & 0.4930 & 0.4178 \\ \hline
FM+LCH & 0.9178 & 0.6820 & 0.7700 & 0.6123 & 0.1422 & 0.4593 & 0.4198 \\
FM+LHS & 0.9180 & 0.6888 & 0.7704 & 0.6278 & 0.1378 & 0.4603 & 0.4164 \\
FM+LHC & 0.9162 & 0.6798 & 0.7698 & 0.6195 & 0.1402 & 0.4608 & 0.4198 \\ \hline
CC+LCH & 0.9504 & 0.7038 & 0.7596 & 0.5864 & 0.1400 & 0.4660 & 0.3869 \\
CC+LHS & 0.9514 & 0.7007 & 0.7626 & 0.5926 & 0.1355 & 0.4793 & 0.3961 \\
CC+LHC & 0.9501 & 0.7072 & 0.7645 & 0.5774 & 0.1349 & 0.4644 & 0.3847 \\ \hline
SVD+LCH & 0.9134 & 0.7076 & 0.7701 & 0.6175 & 0.1240 & 0.4569 & 0.3524 \\
SVD+LHS & 0.9136 & 0.7039 & 0.7722 & 0.6212 & 0.1214 & 0.4630 & 0.3511 \\
SVD+LHC & 0.9126 & 0.7081 & 0.7720 & 0.6133 & 0.1192 & 0.4534 & 0.3602 \\ \hline
NMF+LCH & 0.9522 & 0.7054 & 0.7722 & 0.6233 & 0.1390 & 0.4869 & 0.4030 \\
NMF+LHS & 0.9522 & 0.7026 & 0.7781 & 0.6238 & 0.1466 & 0.4894 & 0.4045 \\
NMF+LHC & 0.9558 & 0.7013 & 0.7692 & 0.6260 & 0.1471 & 0.4852 & 0.4012 \\ \hline
KNN+LCH & 0.9128 & 0.7751 & 0.7659 & 0.6273 & 0.1220 & 0.4365 & 0.3259 \\
KNN+LHS & 0.9133 & 0.7715 & 0.7674 & 0.6287 & 0.1288 & 0.4380 & 0.3388 \\
KNN+LHC & 0.9117 & 0.7753 & 0.7662 & 0.6272 & 0.1327 & 0.4421 & 0.3427 \\ \hline
\end{tabular}
\newline
\caption{Comparison of unexpected recommendations in the Yelp dataset using different geometric structures, ''*'' stands for 95\% statistical significance}
\label{structure1}
\end{table}

\begin{table}
\begin{tabular}{|c|c|c|c|c|c|c|c|c|} \hline
Model & RMSE & MAE & Pre@5 & Rec@5 & Unexp & Ser & Div \\ \hline
NCF+LCH & 0.9635 & 0.7317 & 0.7210 & 0.9795 & 0.0579 & 0.4622 & 0.0478 \\
NCF+LHS & 0.9624 & 0.7310 & 0.7201 & 0.9810 & 0.0586 & 0.4635 & 0.0472 \\
NCF+LHC & 0.9652 & 0.7305 & 0.7214 & 0.9814 & 0.0593 & 0.4647 & 0.0469 \\ \hline
FM+LCH & 1.0275 & 0.7445 & 0.7040 & 0.9656 & 0.0543 & 0.4631 & 0.0393 \\
FM+LHS & 1.0230 & 0.7450 & 0.7031 & 0.9638 & 0.0581 & 0.4637 & 0.0388 \\
FM+LHC & 1.0218 & 0.7472 & 0.7020 & 0.9632 & 0.0561 & 0.4607 & 0.0407 \\ \hline
CC+LCH & 1.0285 & 0.7541 & 0.6865 & 0.9703 & 0.0552 & 0.4619 & 0.0471 \\
CC+LHS & 1.0230 & 0.7539 & 0.6887 & 0.9754 & 0.0587 & 0.4629 & 0.0491 \\
CC+LHC & 1.0200 & 0.7539 & 0.6864 & 0.9730 & 0.0562 & 0.4667 & 0.0498 \\ \hline
SVD+LCH & 0.9937 & 0.7517 & 0.7085 & 0.9594 & 0.0544 & 0.4621 & 0.0499 \\ 
SVD+LHS & 0.9908 & 0.7519 & 0.7093 & 0.9569 & 0.0585 & 0.4614 & 0.0477 \\
SVD+LHC & 0.9884 & 0.7541 & 0.7091 & 0.9474 & 0.0562 & 0.4654 & 0.0485 \\ \hline
NMF+LCH & 1.0262 & 0.7533 & 0.6881 & 0.9775 & 0.0544 & 0.4627 & 0.0499 \\
NMF+LHS & 1.0280 & 0.7594 & 0.6864 & 0.9735 & 0.0584 & 0.4629 & 0.0488 \\
NMF+LHC & 1.0265 & 0.7600 & 0.6853 & 0.9711 & 0.0559 & 0.4677 & 0.0504 \\ \hline
KNN+LCH & 1.0001 & 0.7483 & 0.6907 & 0.9763 & 0.0543 & 0.4631 & 0.0492 \\
KNN+LHS & 0.9981 & 0.7493 & 0.6909 & 0.9743 & 0.0588 & 0.4625 & 0.0488 \\
KNN+LHC & 0.9950 & 0.7524 & 0.6927 & 0.9701 & 0.0564 & 0.4671 & 0.0500 \\ \hline
\end{tabular}
\newline
\caption{Comparison of unexpected recommendations in the TripAdvisor dataset using different geometric structures, ''*'' stands for 95\% statistical significance}
\label{structure2}
\end{table}

\begin{table}
\begin{tabular}{|c|c|c|c|c|c|c|c|c|} \hline
Model & RMSE & MAE & Pre@5 & Rec@5 & Unexp & Ser & Div \\ \hline
NCF+LCH & 0.3799 & 0.2870 & 0.2572 & 0.3638 & 0.7049 & 0.9819 & 0.2538 \\
NCF+LHS & 0.3810 & 0.2854 & 0.2560 & 0.3615 & 0.7070 & 0.9830 & 0.2538 \\
NCF+LHC & 0.3817 & 0.2846 & 0.2549 & 0.3632 & 0.7101 & 0.9852 & 0.2536 \\ \hline
FM+LCH & 0.3906 & 0.2998 & 0.2510 & 0.3278 & 0.7112 & 0.9840 & 0.2518 \\
FM+LHS & 0.3924 & 0.3044 & 0.2498 & 0.3265 & 0.7096 & 0.9833 & 0.2510 \\
FM+LHC & 0.3940 & 0.3056 & 0.2506 & 0.3302 & 0.7177 & 0.9833 & 0.2518 \\ \hline
CC+LCH & 0.4157 & 0.3240 & 0.2564 & 0.3624 & 0.7101 & 0.9817 & 0.2512 \\
CC+LHS & 0.4167 & 0.3296 & 0.2569 & 0.3676 & 0.7053 & 0.9815 & 0.2519 \\
CC+LHC & 0.4151 & 0.3358 & 0.2553 & 0.3659 & 0.7065 & 0.9830 & 0.2508 \\ \hline
SVD+LCH & 0.3841 & 0.2925 & 0.2400 & 0.3277 & 0.7010 & 0.9844 & 0.2408 \\ 
SVD+LHS & 0.3888 & 0.2862 & 0.2455 & 0.3253 & 0.7018 & 0.9810 & 0.2412 \\
SVD+LHC & 0.3836 & 0.2841 & 0.2433 & 0.3271 & 0.7007 & 0.9812 & 0.2454 \\ \hline
NMF+LCH & 0.4423 & 0.3306 & 0.2380 & 0.3491 & 0.7008 & 0.9799 & 0.2488 \\
NMF+LHS & 0.4405 & 0.3330 & 0.2494 & 0.3439 & 0.6999 & 0.9792 & 0.2450 \\
NMF+LHC & 0.4433 & 0.3387 & 0.2420 & 0.3459 & 0.6961 & 0.9803 & 0.2438 \\ \hline
KNN+LCH & 0.4106 & 0.3107 & 0.2584 & 0.3175 & 0.7007 & 0.9817 & 0.2558 \\
KNN+LHS & 0.4088 & 0.3091 & 0.2608 & 0.3212 & 0.7014 & 0.9814 & 0.2558 \\
KNN+LHC & 0.4069 & 0.3099 & 0.2620 & 0.3248 & 0.7073 & 0.9830 & 0.2519 \\ \hline
\end{tabular}
\newline
\caption{Comparison of unexpected recommendations in the Video dataset using different geometric structures, ''*'' stands for 95\% statistical significance}
\label{structure3}
\end{table}

\subsection{Visualization of Latent Embeddings}
\begin{figure*}[h]
\centering
\includegraphics[width=\textwidth]{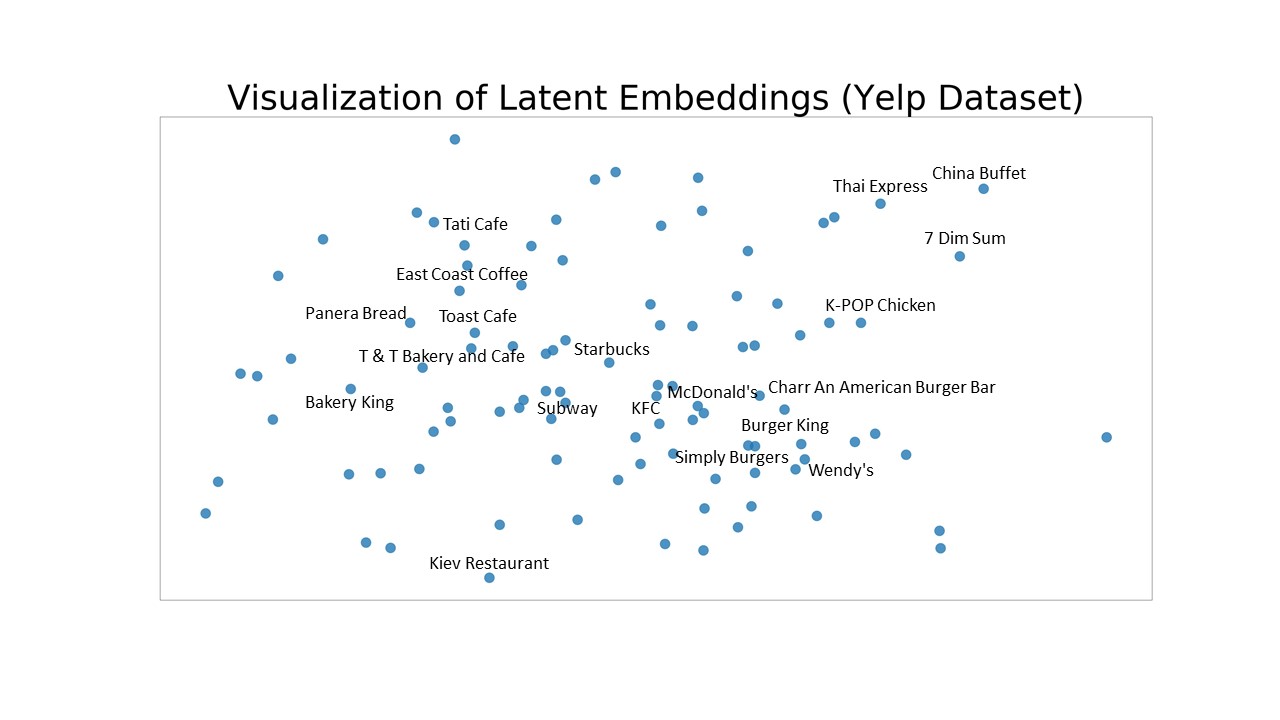}
\caption{t-SNE Visualization of Latent Embeddings}
\label{embedding}
\end{figure*}

Finally, we conduct case study to reveal the effectiveness of modeling unexpectedness through latent embedding approaches. Specifically, we visualize the learned embedding vectors to provide insights of their semantic information in the latent space. Taking the Yelp dataset as an example, we randomly select 100 restaurants from the dataset and obtain their corresponding embeddings through the HINE method. In Figure \ref{embedding}, we show the visualization of those embeddings through t-SNE \cite{maaten2008visualizing}, in which similar restaurants are clustered close to each other. We could see that cafes and bakeries are clustered to the left side, whereas burger bars and fast food restaurants are clustered to the right side, and Asian restaurants are clustered to the far right in the latent space. Therefore, we show that the latent embedding approaches we use in this paper are indeed capable of capturing latent relations among items and thus providing precise modeling of unexpectedness.

\section{Conclusion}
In this paper, we propose novel latent modeling of unexpectedness that simultaneously provides unexpected and satisfying recommendations. Specifically, we define unexpectedness of a new item as the distance between the embedding of that item in the latent space and the closure of all the previously consumed item embeddings. This new definition enables us to capture latent, complex and heterogeneous relationships between users and items that significantly improves performance and practicability of unexpected recommendations. To achieve this, we design a hybrid utility function as the linear combination of estimated ratings and unexpectedness to optimize accuracy and unexpectedness objectives of recommendations simultaneously. Furthermore, we demonstrate that the proposed approach consistently and significantly outperforms all other baseline models in terms of unexpectedness, serendipity and diversity measures without losing any accuracy performance.

The contributions of this paper are threefold. First, we propose \textit{latent} modeling of unexpectedness. Though it is a common idea to explore latent space for recommendations, it is not obvious how to do it for \textit{unexpected} recommendations, as we have discussed in Section 3. Second, we construct the hybrid utility function that combines the proposed unexpectedness measure with the rating estimation value and provides unexpected recommendations based on the hybrid utility values. We demonstrate that this approach significantly outperforms all other unexpected recommendation baselines. Third, we conduct extensive experiments in multiple settings and show that it is indeed the latent modeling of unexpectedness that leads to the significant increase in unexpectedness measures without sacrificing any performance accuracy. Thus, the proposed approach helps users to break out of their filter bubbles.

As the future work, we plan to conduct live experiments within real business environments in order to further evaluate the effectiveness of unexpected recommendations and analyze both qualitative and quantitative aspects in online retail settings through A/B tests. Specifically, we plan to launch our model in an industrial platform and measure its performance using business metrics, including CTR and GMV. Moreover, we will further explore the impact of unexpected recommendations on user satisfaction. Finally, we plan to design algorithms that automatically incorporate the concept of unexpectedness into the deep-learning recommendation framework that optimizes the recommendation performance and the construction of latent embeddings at the same time.

\bibliographystyle{ACM-Reference-Format}
\bibliography{sigproc}

\end{document}